\documentclass[12pt,a4paper,tightenlines,nofootinbib]{revtex4}

\usepackage{amssymb}
\usepackage{amsmath}
\usepackage{epsf}
\usepackage{psfrag}

\def\nablaslash{\not{\hbox{\kern-3pt $\nabla$}}}

\begin{document}

\author{Jaume Garriga$^{1,2,3}$ and Ariel Megevand$^{1,2}$}
\affiliation{$^1$ Departament de F{\'\i}sica Fonamental,
Universitat de Barcelona, Diagonal 647, 08028 Barcelona, Spain}
\affiliation{$^2$ IFAE, Campus UAB, 08193 Bellaterra (Barcelona),
Spain}
\affiliation{$^3$ Institute of Cosmology, Department of Physics and Astronomy, Tufts University, Medford, MA 02155, USA.}
\title{Coincident brane nucleation and the neutralization of $\Lambda$}
\date{\today}

\begin{abstract}

Nucleation of branes by a four-form field has recently been considered in string motivated scenarios for the neutralization of the cosmological constant. An interesting question in this context is whether the nucleation of stacks of coincident branes is possible, and if so, at what rate does it proceed. Feng et al. have suggested that, at high ambient de Sitter temperature, the rate may be strongly enhanced, due to large degeneracy factors associated with the number of light species living on the worldsheet. This might facilitate the quick relaxation from a large effective cosmological constant down to the observed value. Here, we analyse this possibility in some detail.
In four dimensions, and after the moduli are stabilized, branes interact via repulsive long range forces. Because of that, the Coleman-de Luccia (CdL) instanton for coincident brane nucleation may not exist, unless there is some short range interaction which keeps the branes together. If the CdL instanton exists, we find that the degeneracy factor depends only mildly on the ambient de Sitter temperature, and does not switch off even in the case of tunneling from flat space. This would result in catastrophic decay of the present vacuum. If, on the contrary, the CdL instanton does not exist, coindident brane nucleation may still proceed through a ``static" instanton, representing pair creation of critical bubbles -- a process somewhat analogous to thermal activation in flat space. In that case, the branes may stick together due to thermal symmetry restoration, and the pair creation rate depends exponentially on the ambient de Sitter temperature, switching off sharply as the temperature approaches zero. Such static instanton may be well suited for the ``saltatory" relaxation scenario proposed by Feng et al.

\end{abstract}

\maketitle

\section{introduction}

It has long been recognized that the effective cosmological
constant $\Lambda_{eff}$ may have contributions from a four-form
field $F$, and that in such case
\begin{equation}
\Lambda_{eff}=(F^2/2)+ \Lambda \label{1}
\end{equation}
may vary in space and time due to brane nucleation events. This
has led to various proposals for solving the cosmological constant
problem, starting with the pioneering work of Brown and Teitelboim
\cite{teitelboim}. These authors considered a cosmological
scenario where $\Lambda_{eff}$ is initially very large and
positive, due to a large $F^2$ term. The additive constant
$\Lambda$ in (\ref{1}) is assumed to be negative, but not
fine-tuned in any way, so its absolute value is expected to be of
the order of some cut-off scale to the fourth power. During the
cosmological evolution, $\Lambda_{eff}$ is ``neutralized" through
successive nucleation of closed 2-branes (charged with respect to
the form field), which decrease the value of $F$, until eventually
$\Lambda_{eff}$ is relaxed down to the small observed value,
$\Lambda_{obs}$.

One problem with the original scenario is that neutralization must
proceed in very small steps, so that any initially large $\Lambda_{eff}$ can be brought to $\Lambda_{obs}$ without
overshooting into negative values. For that, the charge of the branes
should be tiny, ensuring that $\Delta \Lambda_{eff}\lesssim
\Lambda_{obs}$ at each step. Also, the nucleation rate must be very small, or else the present vacuum would quickly decay. These two constraints make the relaxation process extremely slow on a cosmological time-scale. Meanwhile, ordinary matter in the universe is exponentially diluted by the quasi-de Sitter expansion, resulting in a disappointing empty universe.

Recently, Feng et al. (FMSW) \cite{FMSW} have suggested that
nucleation of coincident branes may offer a solution to the
``empty universe" problem. Their proposal can be summarized as
follows. In the context of M-theory, a stack of $k$ coincident
D-branes supports a number of low energy degrees of freedom,
corresponding to a $U(k)$ super Yang-Mills (SYM) theory living on
the world-sheet. Consequently, the nucleation rate of coincident
branes should be accompanied by large degeneracy factors, and
could in principle be enhanced with respect to the nucleation of
single branes. The charge of a stack of branes can be very large
even if the individual charges are small, facilitating
quick jumps from $\Lambda_{eff}$ to $\Lambda_{obs}$. In this way,
neutralization might proceed very rapidly, perhaps in just a few
``multiple" steps of the right size. Finally, the stability of the
present vacuum could be due to gravitational suppression of the
nucleation rate \cite{deLuccia,teitelboim}.

FMSW argued, rather heuristically, that the nucleation rate of
coincident branes should be enhanced by a factor of the form
\begin{equation}
D \sim e^S, \label{entropyenhancement}
\end{equation}
where $S$ is the entropy of the worldsheet SYM fields. This
entropy was estimated through simple thermodynamic
arguments, as
\begin{equation}
S\sim g_* R^2 T^2,
\label{entro}
\end{equation}
where $g_*$ is the effective number of worldsheet field degrees of freedom, and $R$ is the size of the brane at the time
of nucleation. However it remained unclear in \cite{FMSW} which
temperature $T$ should be used for the worldsheet degrees of
freedom. Brane nucleation takes place in an ambient de Sitter (dS)
space characterized by a Gibbons-Hawking temperature $T_o\propto
\Lambda_{eff}^{1/2}$. The region inside the closed brane has a smaller value of the effective cosmological constant, and is
therefore characterized by a smaller
temperature $T_i$. Feng et al. considered two alternative
possibilities for the temperature of the worldsheet degrees of
freedom: $T_1\sim T_o$ and $T_2\sim (T_o T_i)^{1/2}$. The proposed
enhancement of the nucleation rate and the resulting cosmological
scenarios are quite different in both cases, and therefore it
seems important to try and clarify the issue of which temperature
is the relevant one.

The purpose of this paper is to present a more formal derivation
of the nucleation rate corresponding to multiple brane nucleation.
As we shall see, the temperature relevant for the worldsheet
degrees of freedom is in fact determined by the internal geometry
of the worldsheet \cite{solutions}. For the Coleman-de Luccia (CdL)
instanton, this worldsheet is a 2+1 dimensional de
Sitter space of radius $R$, and the corresponding temperature is
$T\sim R^{-1}$. When substituted into the naive
expression (\ref{entro}), this leads to $S\sim g_*$, independent
of $R$ (and hence on the ambient dS temperatures). As we shall see,
the actual result has a certain dependence on $R$ due
to the anomalous infrared behaviour of light fields in the lower
dimensional de Sitter space, but this results only in a rather mild dependence on the ambient dS temperatures.

We shall also see that de Sitter space allows for a ``static" instanton which may be quite relevant to the nucleation of coincident branes. This is analogous to the instanton for thermal activation in flat space. It has a higher Euclidean action than the CdL solution, and hence (ignoring the degeneracy factor) it seems to represent a subdominant channel of decay. However, we shall argue that, depending on the short distance behaviour of the interactions amongst the branes, the CdL instanton for coincident brane nucleation may simply not exist, and in this situation the static instanton may be the relevant one.
In several respects, the static instanton appears to be better suited to the neutralization scenario proposed by Feng et al. than the CdL one.

The paper is organized as follows. In Section II we
review different proposals for neutralization of $\Lambda_{eff}$ via brane nucleation. Section III contains a discussion of coincident branes in 4 spacetime dimensions. These are obtained from dimensional reduction  of type IIA supergravity in ten dimensions.
In the 4 dimensional picture, the gravitational and four-form forces are both repulsive. However, the two are exactly balanced by the attractive force mediated by the scalar dilaton. In Section IV and V we discuss the stabilization of the dilaton, which is required in a more realistic scenario. After the dilaton aquires a mass, the remaining long range forces are repulsive, rendering the stack of coincident branes unstable, or metastable at best. This has important implications, since the instanton for nucleation of coincident branes will only exist provided that some mechanism causes an attractive interbrane force at short distances.

Section VI contains a description of the CdL instanton for nucleation of coincident branes, highlighting a few limiting cases
of interest.  In Section VII we discuss the corresponding degeneracy factor in the nucleation rate, and we show that its dependence on the ambient de Sitter temperatures is rather mild. We also include a heuristic interpretation of this results based on the observation that the relevant temperature for the worldsheet degrees of freedom is determined by the inverse of the radius of the instanton. Implications for the scenario of \cite{FMSW} are briefly discussed.

Section VIII, is devoted to a study of the ``static" instanton, where the worldsheet has the topology $S^2\times S^1$, and where the intrinsic temperature is comparable to $T_o$. In this case, the dependence of the nucleation rate on the ambient dS temperature is exponential. Coincident brane nucleation can be unsuppressed at large $T_o$ but strongly suppressed at present. Our conclusions are summarized in Section IX. Some technical discussions are left to the Appendices.

To conclude this Introduction, a disclaimer may be useful.
For most of the paper, we shall work directly in four dimensions, and our discussion will be certainly less than rigorous from the string theory point of view. In particular, we shall model the degrees of freedom of a stack of $k$ coincident
2-branes by a weakly coupled $U(k)$ gauge theory on the world-sheet. This may or may not correspond to a true dimensional reduction from M-theory, but it should at least represent some of the broad features of the degeneracy factors.

\section{Neutralization $v.s.$ randomization}

In four dimensions, a four form can be written as ${\cal
F}_{\mu\nu\rho\sigma} = F \sqrt{-g}\epsilon_{\mu\nu\rho\sigma}$
where $g$ is the determinant of the metric and $\epsilon$ is the
Levi-Civita symbol. This field has no propagating degrees of
freedom, since in the absence of sources the equation of motion
$d^*{\cal F}=0$ implies that $F$ is a constant. This simply gives
a contribution to the effective cosmological constant
which gets added to the true cosmological constant, or vacuum
energy density $\Lambda$,
\begin{equation}
\Lambda_{eff}={F^2 \over 2}+\Lambda. \label{lambdaeff}
\end{equation}
Four-forms may couple to ``charged" 2-dimensional extended
sources, or 2-branes, through a term of the form
\begin{equation}
q\int_\Sigma {\cal A},
\end{equation}
where the integral is over the worldsheet of the extended object
and ${\cal A}$ is the 3-form potential (${\cal F}=d{\cal A}$). In
this case, $F$ changes by
\begin{equation}
\Delta F = q\label{deltaf}
\end{equation}
accross a brane of charge $q$. Consequently, $F$ can decay through
nucleation of closed spherical branes. The process is analogous to
pair creation in the presence of an electric field, and very
similar to false vacuum decay in field theory \cite{coleman}. The
closed brane is the boundary of a newly formed ``true" vacuum
bubble, where the field strength differs from the original value
by the amount (\ref{deltaf}). After nucleation, the radius of the
bubble grows with constant proper acceleration, and the volume
occupied by the new phase keeps increasing. Further nucleation
events take place in the region with a low $F$, lowering
$\Lambda_{eff}$ even further. In the absence of gravity, this
``neutralization" would proceed as long as $F>q/2$, wiping out any
large initial value of $F^2$ and leaving us with
a large negative cosmological constant $\Lambda^{final}_{eff} \sim \Lambda + O(q^2)$. Of course, this is not what we want.

Gravitational effects can improve the picture dramatically
\cite{teitelboim}. In particular, tunneling from flat or Anti-de
Sitter space (with a negative cosmological constant) is forbidden
provided that the squared tension of the branes is sufficiently
large compared with the jump $\Delta\Lambda_{eff}\sim q F$ in
energy density across the brane,
\begin{equation}
\sigma^2> (4/3) q F M_p^2. \label{stability}
\end{equation}
Here $\sigma$ is the brane tension and $M_p^2=1/(8\pi G)$ is the
square of the reduced Planck mass. This gravitational ``shutdown"
of brane nucleation could be useful, since an initially large
$\Lambda_{eff}$ may eventually get ``neutralized" to a value which
is much smaller than $\Lambda$ in absolute value
\cite{teitelboim}. Suppose that the true cosmological constant is
negative $\Lambda<0$. As long as $\Lambda_{eff}>0$, branes keep
nucleating. But once a vacuum with $\Lambda_{eff}\leq 0$ is
reached, the process stops provided that Eq. (\ref{stability}) is
satisfied. After that, the vacuum becomes absolutely stable. Brown
and Teitleboim conjectured that we may live in one such vacuum,
where the effective cosmological constant is expected to be of the
order of the energy density gap between neighboring vacua
\begin{equation}
|\Lambda^{final}_{eff}| \sim F \Delta F \sim q |\Lambda|^{1/2}.
\label{geo}
\end{equation}
In this vacuum, $|F| \sim |\Lambda|^{1/2}$, and the huge negative
bare cosmological constant is almost completely cancelled by the
$F^2$ contribution in the final state.

For sufficiently small charge $q$, Eq. (\ref{geo}) leads to a
suppression of $|\Lambda^{final}_{eff}|$ relative to the true
vacuum energy $|\Lambda|$, which might be helpful in solving the
``old'' fine tuning problem of the cosmological constant. Particle
physics models suggest $|\Lambda| \gtrsim (TeV)^4$. Hence, we need
\begin{equation}
q\lesssim 10^{-35} (eV)^2, \label{smallcharge}
\end{equation}
so that the final value $|\Lambda^{final}_{eff}|$ is consistent
with the observed value $|\Lambda^{obs}_{eff}| \sim
10^{-11}(eV)^4$. The constraint (\ref{smallcharge}) seems rather
demanding, since in the context of supergravity we would expect
$q$ to be closer to the Planck scale. This is the so-called ``gap
problem".
FMSW argued that the smallness of the charge could be due to the
wrapping of branes on degenerating cycles in the extra dimensions.
A successful implementation of this idea has not yet been
presented, but some plausibility arguments have been
given in \cite{FMSW}.
Alternatively, in a different context, it has been suggested that
branes with a very tiny charge $q$ may arise due to symmetries of
the theory. An explicit example was given in \cite{alexgia}, where
the branes are not fundamental objects but domain walls of a
broken discrete symmetry. This same symmetry suppresses the
coupling of the domain walls to the four-form $F$ without any
fine-tuning of parameters (see also \cite{alexgiahierarchy} for a
fuller discussion).

A more severe problem of the Brown and Teitelboim neutralization
scenario, is the ``empty universe'' problem which we discussed in
the Introduction. By combining the condition (\ref{smallcharge})
with the stability condition (\ref{stability}), it can be shown
that the time required to reach the value (\ref{geo}) is huge
compared with the age of the universe \cite{teitelboim}. By the
time the effective cosmological constant would be wiped out, all
other forms of matter would have also been diluted exponentially,
in clear contradiction with observations. Furthermore, the
endpoint of neutralization would be a state with vanishing or
negative effective cosmological constant, whereas the observed
value $\Lambda_{eff}^{obs}$ is positive.

One way around the empty universe problem is to consider a
slightly different scenario, where the effective cosmological
constant is ``randomized" (rather than neutralized) during
inflation. Assume, for simplicity, that the energy scale of
inflation is much larger than $\Delta\Lambda_{eff}$. In an inflationary phase,
brane nucleation
processes may increase as well as decrease the value of $F^2$
\cite{solutions} . Thus, $\Lambda_{eff}$ will randomly
fluctuate up and down the ladder as a result of brane nucleation.
Inflation is generically eternal to the future and there is an
unlimited amount of time available for the randomization process
to take place before thermalization \cite{solutions}. Assume that
the tunneling barriers are sufficiently high, so that no
nucleation events happen in the last 60 e-foldings of inflation,
or during the hot phase after thermalization up to the present
time. In this scenario there is no empty universe problem: the
local value of $\Lambda_{eff}$ is decided many e-foldings before
the end of inflation, and a wide range of values of $\Lambda_{eff}$ will be found in
distant regions of the universe, separated from each other by
distances much larger than the present Hubble radius. Some of
these regions will just happen to have a very tiny
$\Lambda_{eff}$. In combination with anthropic selection effects,
this approach may be used to explain the smallness of the observed
effective cosmological constant
\cite{MSW,solutions,alexgia,prelam}. This may also explain the
so-called cosmic time coincidence, or why do we happen to live at
the time when an effective cosmological constant starts dominating
\cite{mario,solutions,prelam}.

Bousso and Polchinski \cite{bopo} have proposed a somewhat related
``randomization" scenario which, moreover, does not rely on branes
with the exceedingly small charge $q$ satisfying
(\ref{smallcharge}). In the context of string theory one may
expect not just one but many different four-form fluxes $F_i$
coupled to branes with different charges $q_i$. Each one of these
fluxes is quantized in units of the charge, so that $F_i = n_i
q_i$. In this case, the condition for a generic negative
cosmological constant to be compensated for by the fluxes is
$|\Lambda_{eff}|=|\sum_{i=1}^J (q^2_i n_i^2)/2 + \Lambda|\lesssim
\Lambda_{obs}$. The larger the number of different fluxes, the
denser is the discretuum of possible values of $\Lambda_{eff}$,
and the easier it is to find a set of values of $n_i$ such that
the above inequality is satisfied. A sufficiently dense discretuum
is typically obtained provided that the number of fluxes is
sufficiently large $J\sim 100$, even if the individual charges
$q_i$ are Planckian (a smaller number of fluxes $J\gtrsim 6$ may
be enough in a scenario with large extra dimensions, where the
charges $q_i$ are suppressed with respect to the Planck scale by a
large internal volume effect). In the Bousso-Polchinski model,
$\Lambda_{eff}$ is typically very large, and drives an exponential
expansion at a very high energy scale. Suppose we start with a
single exponentially expanding domain characterized by a set of
integers $\{n_i\}$. Whenever a brane of type $j$ nucleates in this
region, the integer $n_j$ will change by one unit inside of the
brane. The newly formed region will itself expand exponentially,
creating a huge new domain. The nucleation of further branes
within this region will cause an endless random walk of the values
of $\{n_i\}$, which will sample the whole discretuum of values of
$\Lambda_{eff}$. Eventually, a bubble will nucleate where
$\Lambda_{eff}$ is comparable to the observed value. The
nucleation of this last bubble is still a high energy process,
which kicks some inflaton field off its minimum, and starts a
short period of inflation within this last bubble. This period of
``ordinary" inflation is necessary in order to produce the entropy
we observe, thereby avoiding the empty universe problem. Of
course, some anthropic input is still necessary in this approach
in order to explain why, out of the discretuum of possibilities,
we live in a vacuum with small $\Lambda_{eff}$.

Nucleation of coincident branes would drastically modify the
neutralization scenario of Brown and Teitelboim, as well as the
randomization scenarios sketched above. As proposed in
\cite{FMSW}, in the case of neutralization, this modification may
lead to a solution of the empty universe problem. In the
randomization scenarios there is no such problem, and it is
unclear whether an enhancement of the multiple brane nucleation
rate is desirable at all. This enhancement would trigger large
jumps in the effective cosmological constant, making the
calculation of its spatial distribution more complicated than for
single brane nucleation. Thus, it is of interest to understand the
conditions under which multiple brane nucleation is allowed, and
what are the degeneracy factors which might enhance their
nucleation rate relative to the nucleation of single branes.
Before addressing this issue, it will be convenient to present a
short discussion of coincident branes.

\section{Coincident Branes in 4D}

In the context of string theory, one may consider stacks of $k$
coincident D-p-branes. Each brane in the stack has charge $q$ with
respect to the form field ${\cal A}$. Thus, in four noncompact
dimensions, a pair of parallel 2-branes repel each other with a
constant force per unit area given by
\begin{equation}
f_{q}=q^2/2, \label{qfield}
\end{equation}
due to the four-form field interaction.
In the ten dimensional theory, the repulsive force due to ${\cal
A}$ is balanced with other contributions from the closed string
sector, such as the graviton and dilaton. As a result, there are
no net forces amongst the different branes on the stack.

It is in principle possible to maintain this delicate balance by
suitable compactification from 10 to 4 dimensions. In 4D, the
branes look like domain walls, and their interaction through the
ordinary graviton leads to a mutual repulsive force given by
\cite{vilenkin81},
\begin{equation}
f_{\sigma}=3 \sigma^2/4 M_p^2. \label{dwfield}
\end{equation}
Hence, both forces given by (\ref{qfield}) and (\ref{dwfield})
tend to push the branes apart from each other. On the other hand,
some of the higher dimensional gravitational degrees of freedom
are represented by scalars in 4D, and these, together with the
dilaton, lead to attractive forces.

\subsection{Dimensional reduction}

Let us first consider the case of D-2-branes in 10 dimensional
Type IIA supergravity. The relevant part
of the action is given by (see e.g. \cite{johnson})
\begin{equation}
S_{10} = {M_{10}^8\over 2}\int d^{10}x \sqrt{G}\left[
e^{-2\phi}\left(R+4(\nabla \phi)^2\right)- {1\over 2\cdot 4!}
\hat{\cal F}^2\right] - T_2\int_\Sigma d^3\xi \sqrt{G_\Sigma}\
e^{-\phi} + \hat q_2 \int_\Sigma d^3\xi\ \hat{\cal A}.
\end{equation}
Here, $\sqrt{G}$ and $\sqrt{G_\Sigma}$ are the determinants of the
10 dimensional metric $G_{AB}$ and of the metric induced on the
worldsheet $\Sigma$, respectively, whereas $R$ is the Ricci scalar
corresponding to $G_{AB}$. The hats on the four-form, the gauge
potential, and the corresponding charge, are introduced in order
to distinguish them from the four dimensional ones which will be
used below, and which differ from those by constant normalization
factors. Compactifying on a Calabi-Yau manifold $K$ through the
ansatz
$$
ds_{10}^2= e^{2\phi-6\psi} g_{\mu\nu} dx^{\mu}dx^{\nu} + e^{2\psi}
d K_6^2,
$$
where the Greek indices run from 0 to 3, we readily obtain the
following four-dimensional action:
\begin{equation}
S_4 = {M_p^2\over 2}\int d^4 x \sqrt{g}\left\{{\cal R} -{2\over 7}
(\partial \hat\varphi)^2-{6\over 7}(\partial\hat\sigma)^2 -
{1\over 2\cdot 4!}\ e^{-2\hat{\varphi}} \hat{\cal F}^2 \right\}
\label{reduced4d}
\end{equation}
$$
- T_2 \int_{\Sigma} d^3 \xi \sqrt{\gamma}\ e^{\hat {\varphi}} +
\hat q_2 \int_{\Sigma} \hat{\cal A}.
$$
Here, $M_p^2= M_{10}^8 V_6$, where $V_6$ is the coordinate volume
of the manifold $K$, ${\cal R}$ is the Ricci scalar for the metric
$g$, and we have introduced two linear combinations of the
internal volume modulus $\psi$ and the dilaton $\phi$
$$
\hat\varphi = 2\phi - 9\psi, \quad\quad \hat\sigma = \phi-\psi.
$$
The field $\hat\sigma$ decouples from the branes, and shall be
ignored in what follows. In Ref.\cite{FMSW}, a different expression
was given for the dimensionally reduced action, because no modulus
was introduced for the size of the internal space. However, as we
shall see, such modulus is necessary for the cancellation of forces
among the branes.

A more direct route to the 4-dimensional theory (\ref{reduced4d}),
which will be slightly more convenient for our discussion, is to
start directly from 11-dimensional supergravity and compactifying
on a 7 dimensional internal space. The action in eleven dimensions
is given by
\begin{equation}
S_{11} = {M_{11}^9\over 2}\int d^{11}x \sqrt{G}\left[R- {1\over
2\cdot 4!} \hat{\cal F}^2\right] - T_2\int_\Sigma d^3\xi
\sqrt{G_\Sigma}+ \hat q_2 \int_\Sigma \ \hat{\cal A}.
\end{equation}
where $T_2=\hat q_2= 2\pi M_{11}^3$. Introducing the ansatz
\begin{equation}
ds_{11}^2= e^{-7\hat\psi} g_{\mu\nu} dx^{\mu}dx^{\nu} +
e^{2\hat\psi} d \Omega_7^2, \label{ansatz11}
\end{equation}
with a Ricci flat internal manifold, we find
\begin{equation}
S_4 = {M_p^2\over 2}\int d^4 x \sqrt{g}\left\{{\cal R} -{2\over 7}
(\partial \hat\varphi)^2-{1\over 2\cdot 4!}\ e^{-2\hat{\varphi}}
\hat{\cal F}^2 \right\} + \cdots, \label{reduced4d2}
\end{equation}
where $M_p^2= M_{11}^9 V_7$ and
$$
\hat\varphi=-{21\over 2}\hat\psi.
$$
Not surprisingly, this has the same form as (\ref{reduced4d}),
since after all the 10-dimensional Lagrangian can be obtained from
the 11-dimensional one by compactifying on a circle.

Eq. (\ref{reduced4d2}) is a particular case of the slightly more
general action in four dimensions:
\begin{equation}
S_4= {1\over 2}\int d^4 x \sqrt{g}\left\{M_p^2[{\cal R} -
(\partial \varphi)^2] - {1\over 4!}\ e^{-2\alpha{\varphi}} {\cal
F}^2 \right\}- \sigma \int_{\Sigma} d^3 \xi \sqrt{\gamma}\
e^{\alpha {\varphi}} + q \int_{\Sigma} {\cal A} . \label{action4d}
\end{equation}
The parameter $\alpha$ characterizes the scalar charge of the
brane. As we shall see below, linearizing in $\varphi$ around
$\varphi=0$, it can be easily shown that the scalar force is given
by
\begin{equation}
f_{e}= - e^2/2,\label{scalarforce}
\end{equation}
where we have introduced the scalar charge
\begin{equation}
e \equiv \alpha {\sigma/M_p}. \label{alpha}
\end{equation}
Thus, from  (\ref{qfield}), (\ref{dwfield}) and
(\ref{scalarforce}), the branes will be in indifferent equilibrium
provided that the following relation holds:
\begin{equation}
Q^2\equiv \left[{e^2 - q^2\over 2} - {3 \sigma^2 \over 4
M_p^2}\right]= 0.\label{Q2}
\end{equation}
From (\ref{alpha}), this condition can be rewritten as
\begin{equation}
q =\left(\alpha^2 - {3\over 2}\right)^{1/2} {\sigma\over
M_p}.\label{bog}
\end{equation}
Note that the case of Type IIA supergravity discussed above,
corresponds to $\alpha=\sqrt{7/2}$, $q=(\sqrt{2}/M_p)\ \hat q_2$,
and $\sigma= T_2$ [note that $\hat {\cal A} =
({\sqrt{2}/M_p}){\cal A}$]. Therefore (\ref{bog}) is satisfied
provided that $T_2=\hat q_2$, the usual BPS condition.

\subsection{Multiple brane solutions}

\begin{figure}[tbh]
\psfrag{Fq}[][r]{$F=q$} \psfrag{Fmq}[][l]{$F=-q$}
\psfrag{F0}[][]{$F=0$}\psfrag{exp}{$e^{-\alpha\varphi}$}
\psfrag{d}{$d$}\centering \epsfysize=5cm \leavevmode
\epsfbox{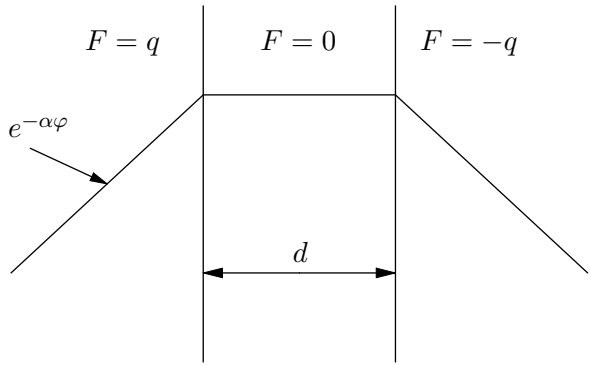}
 \caption{Configuration of two parallel branes} \label{fibranas}
\end{figure}

Multiple brane solutions to (\ref{action4d}) can easily be
constructed provided that (\ref{bog}) is satisfied. In the bulk,
the equation of motion for ${\cal A}$ leads to
$$
{\cal F}_{\alpha\beta\gamma\delta}= F\  e^{2\alpha\varphi}
\sqrt{g}\ \epsilon_{\alpha\beta\gamma\delta},
$$
where $F$ is a constant and $\epsilon$ is the Levi-Civita symbol.
At the branes, this constant jumps by the amount
$$
\Delta F = q
$$
It can be checked that the remaining equations of motion for the
scalar and the gravitational field follow from the action
\begin{equation}
S_4= \sum_i \int_i d^4 x\sqrt{g} \left\{{M_p^2\over 2}[{\cal R} -
(\partial \varphi)^2] -  V_i(\varphi) \right\} - \sigma
\int_{\Sigma} d^3 \xi \sqrt{\gamma}\ e^{\alpha {\varphi}},
\label{action4dII}
\end{equation}
where
\begin{equation}
V_i(\varphi)={F_i^2\over 2} \ e^{2\alpha{\varphi}},
\label{run}
\end{equation}
and the sum is over the regions with different values of $F$. With
the metric ansatz
$$
ds^2= w^2(z) \eta_{ab}d x^{a}dx^{b} + d z^2,
$$
where Latin indices run from 0 to 2, the solution is given by
\begin{equation}
e^{-\alpha \varphi} = c_i \pm {\alpha^2 \over
(\alpha^2-{3/2})^{1/2}}{|F_i|\over M_p}\ z,
  \label{branesolution}
\end{equation}
and
\begin{equation}
w(z) = e^{-{\varphi\over 2 \alpha}}. \label{solutiona}
\end{equation}
Here, $c_i$ are integration constants. These, and the sign option
in (\ref{branesolution}), must be chosen so that the junction
conditions for the gravitational field and for the scalar field
are satisfied at the branes. For the gravity part, the condition
is \cite{israel},
\begin{equation}
\left[ K_{ab}\right] =-4\pi G\sigma\ e^{\alpha {\varphi}} \gamma
_{ab},  \label{discont}
\end{equation}
where $\left[ K_{ab}\right] $ is the difference of extrinsic
curvature on the two sides of the brane and $\gamma _{ab}$ is the
world-sheet metric. In the present case, this reduces to
\begin{equation}
\left[{w'\over w}\right] = -{\sigma\over 2 M^2_p}\ e^{\alpha
{\varphi}}. \label{junctiona}
\end{equation}
For the scalar field, the junction condition which follows from
the equation of motion at the brane reduces to
\begin{equation}
\left[\varphi'\right]= \alpha{\sigma\over M_p^2}\ e^{\alpha
{\varphi}}, \label{junctionphi}
\end{equation}
which is consistent with (\ref{junctiona}) and (\ref{solutiona}).
For instance, a solution with two branes separated by a distance
$d$ is given by flat space ($w=1,\varphi=0$) in the region between
the branes, and by
\begin{equation}
e^{-\alpha\varphi}=1 - \alpha^2 {\sigma\over M^2_p}\ (|z|-d/2)
\quad\quad (|z|> d/2) \label{sandwich}
\end{equation}
in the exterior region (see Fig \ref{fibranas}). The solution corresponds
to two branes of charge $q$ interpolating between regions with
$F=q$ and $F=-q$, separated by a region with $F=0$.

Note that the solution (\ref{sandwich}) contains a singularity at
a distance $z\sim M_p^2/\alpha^2\sigma$ (where the warp factor $w$
and the volume of the internal space vanish). In what follows,
however, we shall not be interested in flat, infinite
branes, such as the ones discussed above. Rather, we shall be
interested in compact instanton solutions of finite size, in a
theory with a stabilized dilaton. In this case, the singularities
due to linear potentials of the form (\ref{sandwich}) should not
arise, but the solutions discussed above should remain a good
description in the vicinity of the branes. Another consequence of
stabilizing the dilaton is that the perfect balance of forces
amongst the branes will be spoiled at distances larger than the
inverse mass of the moduli, as we now discuss.

\section{Stabilizing the moduli}

As it stands, the dimensionally reduced supergravity lagrangian
(\ref{reduced4d}) [or its generalization (\ref{action4d})] is not
useful for discussing the neutralization of the cosmological
constant. The lagrangian does not include the bare
cosmological term $\Lambda$, which is precisely the subject of our
interest, and the term proportional to $F^2$ does not behave
as an effective cosmological constant, but rather as an
exponential potential for a modulus $\varphi$ (which is not flat
enough to mimick the vacuum energy \footnote{Exponential potentials
such as the one appearing in (\ref{action4dII}) have been
thoroughly studied in the literature, and it is known that they
can drive cosmological solutions with a power-law scale factor
\cite{halliwell87}. Such attractor solutions are approached for a
wide range of initial conditions, and the resulting expansion can
be accelerating or decelerating, depending on whether
$\alpha^2<1/2$ or $\alpha^2>1/2$. For $\alpha^2 < 3/2$ the
cosmological scale factor approaches $a(t)\sim t^{1/2\alpha^2}$,
where $t$ is cosmological time. This solution corresponds to an
effective equation of state $p=[(4\alpha^2/3)-1]\rho$, where the
ratio of kinetic and potential energies of the scalar field
$\varphi$ remains constant. In our case, from (\ref{bog}), we need
$\alpha^2> 3/2$ and therefore the kinetic term becomes completely
dominant in the long run, which leads to $p=+\rho$. Hence, by
itself, the $F^2$ term does not behave like an effective
cosmological constant.}). A more realistic model is obtained by introducing a stabilization mechanism which fixes the
expectation value of $\varphi$, and gives it some mass $m$.
Once $\varphi$ is stabilized the $F^2$ term does behave as a contribution to $\Lambda_{eff}$.
Stabilization is also desirable because the dilaton and the radion
moduli (corresponding to the size of extra dimensions) mediate
scalar interactions of gravitational strength, which are severely
constrained by observations.
The study of mechanisms for
stabilization is currently an active topic of research (see e.g. \cite{modstab,Giddings:2003zw} and references therein).
Although the details of stabilization will not be too
important in our subsequent discussion, it may be nevertheless illustrative to have in mind a specific toy mechanism (for which we do not claim any rigorous justification in the context of string theory).

The general problem is that the potential (\ref{run}) has no minimum and
leads to a runaway dilaton. In order to create a minimum, let us consider
two contributions which may be added to (\ref{run}).
First of all, instead of using a Ricci flat internal
manifold in (\ref{ansatz11}) we may compactify on an Einstein
manifold, with
\begin{equation}
R_{ab}^{\Omega}= 6 K g_{ab}^{\Omega}.\label{einstein}
\end{equation}
Upon dimensional reduction, the curvature $K$ contributes an
exponential term to the effective potential for
$\varphi$. However, this will still not be sufficient for stabilizing
the internal volume in an interesting way. In fact, as shown by
Nunez and Maldacena \cite{nogo}, there are no static
compactifications of the classical supergravity Lagrangian with a
positive effective four-dimensional cosmological constant
$\Lambda_{eff}> 0$, and so far we have added nothing to the classical Lagrangian.
Thus, in order to implement the four
dimensional situation of our interest, a third term related to
quantum corrections has to be considered. Following Candelas and
Weinberg \cite{cawe}, we may consider the Casimir energy of bulk
fields (we are of course assuming that supersymmetry is broken, so that the Casimir contributions of bosons and fermions do not cancel each other exactly).

In the example considered by Candelas and Weinberg \cite{cawe},
besides the
Casimir energy term, a higher dimensional
cosmological term $\Lambda_{4+n}$ was used, and the internal manifold was taken to be a space of constant
positive curvature ($K=1$). In 11-D supergravity a cosmological constant
is not allowed, so here we use the $F^2$ flux instead. Also, we will need to compactify on a negatively curved internal
manifold ($K=-1$), or else the dilaton would be stabilized at
negative $\Lambda_{eff}$.

The versatility of negatively curved
compactifications has been stressed in \cite{nemanja}. In particular, they have
the interesting property of rigidity, which means they do not lead to other
moduli besides the size of the internal space. Compact
hyperbolic manifolds (CHM) can be obtained from the maximally
symmetric negatively curved space $H_7$ through identifications by a discrete isometry group $\Gamma$. The volume of
$H_7/\Gamma$ is given by
\begin{equation}
V_7= r_c^7 e^\gamma, \label{v7}
\end{equation}
where $r_c$ is the curvature radius of the manifold, related to
the curvature parameter in (\ref{einstein}) by
$
K= -{1/r_c^2}.
$
The factor $e^\gamma$ depends on the topology, and it is bounded
below but not above. If $L$ is the largest distance around the
manifold, then for $L\gg r_c/2$ we have $e^\gamma\sim e^{6L/r_c}$.
The Kaluza Klein (KK) spectrum in this manifold is believed to have a
mass gap, bounded below by
$
m_{KK} \sim e^{-\hat\psi} r_c^{-1}.
$
From the 11 dimensional point of view the Casimir energy density
scales like
\begin{equation}
\rho_C^{(11)}= C m_{KK}^{11}= C e^{-11\hat\psi} r_c^{-11}.
\label{rho11}
\end{equation}
The factor $C$ can be
estimated by naive dimensional analysis as
$
C\sim \beta \nu,
$
where $\nu\sim 10^{3}$ is the number of physical polarizations of
bulk fields, and $\beta$ is some small one-loop factor. This factor
will depend on the precise topology of the compact hyperbolic
manifold, but it could plausibly be in the range $\beta\sim
10^{-2}-10^{-4}$. Hence, the parameter $C$ could be of order one.
Explicit calculations for different choices of the manifold
$H_7/\Gamma$ have not been performed, and are well beyond the
scope of the present paper (see e.g. \cite{casimir} and references therein). In what
follows we shall leave $C$ unspecified, assuming that a
compactification exists where $|C|\gtrsim 1$.
Multiplying the higher dimensional energy density $\rho_C^{(11)}$ given
in (\ref{rho11}) by the size of the internal manifold, $V_7
e^{7\hat\psi}$, and by a factor $e^{-14\hat\psi}$
which arises from the four-dimensional volume element in going to
the Einstein frame, the effective
potential which appears in Eq. (\ref{action4dII}) gets replaced by
\begin{equation}
V_i(\varphi) =    -21\ {M_{11}^9 V_7\ K} e^{-9\hat\psi}+ {C\ V_7
r_c^{-11}} e^{-18\hat\psi}+ {F_i^2 \over 2}
e^{-21\hat\psi},\label{stabpot}
\end{equation}
where we have also added the curvature contribution.
Here $\hat\psi=-(2/21)\alpha\varphi$, where $\alpha= \sqrt{7/2}$.

We can always adopt the convention that $r_c=1/M_{11}$, since a change of $r_c\to e^{-\lambda} r_c$ can be reabsorbed by a shift in $\hat\psi\to \hat\psi + \lambda$ and a constant re-scaling of the four-dimensional
metric $g_{\mu\nu}\to e^{7\lambda} g_{\mu\nu}$ (in this frame, the curvature of the
manifold is of the order of the higher dimensional Planck scale
for $|\hat\psi| \lesssim 1$). Since $F_i = n_i q$, where
$n_i$ is an integer and $q= (\sqrt{2}/M_p)\hat q_2 = 2\sqrt{2} \pi
M_{11}^3/M_p$, we have
\begin{equation}
V_i(\varphi) = M_p^2M_{11}^2\left[ 21\ e^{-9\hat\psi} + C\
e^{-18\hat\psi}+ 4\pi^2 n_i^2 \left({M_{11}\over M_p}\right)^4\
e^{-21\hat\psi}\right].\label{stabpot2}
\end{equation}
As illustrated by Eq. (\ref{v7}) a large value of $V_7$ in units
of $r_c=1/M_{11}$ can be obtained by using a manifold with
sufficiently complicated topology. Since $M_p^2=M_{11}^9 V_7 =
M_{11}^2 e^\gamma$, the factor $(M_{11}/M_p)^4= e^{-2\gamma}$ in
the last term of (\ref{stabpot2}) can be rather small. The scale
$M_{11}$ could be as low as the $TeV$, in which case that
factor can be as low as $10^{-64}$. Moreover, as emphasized in
\cite{nemanja}, this can be achieved for CHM even if the linear
size $L$ of the internal manifold is not very much larger than
$r_c$. In Fig. 2 we plot the effective potential for $C=-20$,
$e^{-\gamma}=10^{-3}$ and values of $n_i=485,487,489$ and $491$.

When a single brane nucleates, it changes $n_i$ by one unit, and hence
changes the value of the effective potential at the minimum (changing therefore the effective cosmological constant).
If the discretuum of values of $F_i$
were sufficiently dense, then there would always be one of the
minima of the effective potential where the vacuum energy
is sufficiently small to
match observations. In the case we have considered here, the
discretuum is not dense enough. The cancelation between the last
term in (\ref{stabpot}) and the other two requires $n_i \sim
(M_p/M_{11})^2$, and so the gap between levels near $V_i=0$ can be
estimated as $\Delta V_i \sim M_{11}^4$, which is far too large.
This situation can be remedied by considering 5M branes wrapped
around 3-cycles in the internal space. As emphasized by Bousso and
Polchinski \cite{bopo}, these are coupled (with different charges)
to additional fluxes, and a large number of fluxes will result in
a much denser spectrum. Alternatively, FMSW have suggested that
the branes may wrap a degenerating cycle \cite{FMSW}, in which
case the individual charges might themselves be exponentially
smaller than the fundamental scale, resulting also in a
sufficiently dense discretuum.
\begin{figure}[tbh]
\psfrag{V}{$V$} \psfrag{f}{$\hat\psi$}
\centering \epsfysize=5cm \leavevmode \epsfbox{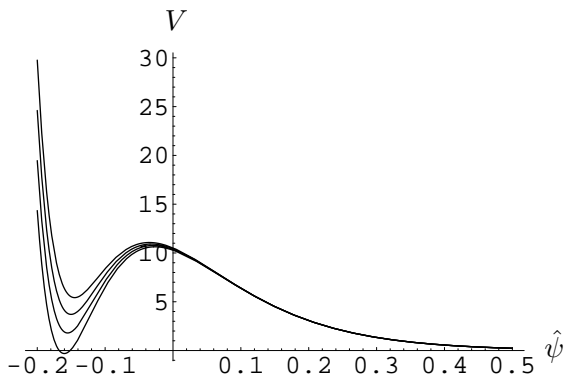}
 \caption{The effective potential (\ref{stabpot2}), for different
 values of the integer $n_i$ which characterizes the quantized flux
 of the four-form $F$.} \label{moduli}
\end{figure}

Before closing this Section, we should note that the location where
the modulus sits is basically determined by the competition between
the  two first terms in (\ref{stabpot}), corresponding to
curvature and Casimir energy. The physical curvature radius of the
internal space is therefore stabilized at $r_{phys}=r_c
e^{\hat\psi}\sim M_{11}^{-1} C^{1/(D-2)}$, where $D=11$ is the spacetime
dimension. Thus, unless the constant $C$ in Eq. (\ref{rho11}) is exceedingly
large, the compactification scale is comparable to the inverse of the higher
dimensional cut-off scale $M_{11}$. In such case the semiclassical analysis
which we have entertained above is not justified, since higher order
corrections will be just as important as the one loop effect which we
have included. This appears to be a generic problem when we try to stabilize
by making the curvature and the Casimir terms comparable, as in the Candelas
and Weinberg example \cite{cawe}. There, the problem was not quite as poignant, since the constant $C$ could be made very large by adding a sufficient number of
fields (also, as it is clear from the above estimate of $r_{phys}$, the problem
is somewhat milder if the number of extra dimensions is smaller). Here we shall
not dwell on this problem, since the main purpose of the above discussion is just
to illustrate the role of the four-form in obtaining a discretuum of states. For more fundamental approaches to moduli stabilization in the present context, the reader is referred to \cite{modstab} and references therein. Our ensuing discussion will be
largely independent of the details of the stabilization mechanism.

\section{Inter-brane forces}

Inevitably, the stabilization of $\varphi$ spoils the perfect
balance of forces. At distances larger than the inverse
of the mass of the modulus $m^{-1}$, the remaining gravitational
and four-form interactions are both repulsive, and lead to a
linear potential per unit surface of the form
\begin{equation}
{\cal V}(d) \approx -\left[{q^2 \over 2} + {3\sigma^2 \over 4
M_p^2} \right]\ d \quad\quad \quad (d\gg m^{-1})
\end{equation}
where $d$ is the distance between the branes.

To investigate the behaviour of the interaction potential at distances
shorter than $m^{-1}$, let us first consider the situation where gravity
and the 3-form gauge potential ${\cal A}$ are ignored, and
the branes interact only through a
scalar field $\varphi$ of mass $m$. The action is given by:
\begin{equation}
S= -{M_p^2\over 2}\int d^4 x \left[
(\partial \varphi)^2 + m^2 \varphi^2 \right]
-\sigma \int_{\Sigma} d^3 \xi \sqrt{\gamma}\ e^{\alpha\varphi}.
\label{actiontoy3}
\end{equation}
The solution with a single brane on the plane $z=0$ has the cusp
profile
\begin{equation}
\varphi = \varphi_0 e^{-m|z|}, \label{cusp}
\end{equation}
where $\varphi_0$ is a solution of
\begin{equation}
\varphi_0\ e^{-\alpha\varphi_0} = - {\alpha \sigma\over 2m M_p^2}
\equiv -{e\over 2 m M_p}. \label{transcendental}
\end{equation}
The energy per unit area of this configuration is given by
\begin{equation}
\sigma_1= \sigma  e^{\alpha \varphi_0} + {M_p^2\over 2}\int dz
\left( \varphi'\ ^2 + m^2 \varphi^2 \right)=\sigma  e^{\alpha
\varphi_0} + m M_p^2 \varphi_0^2.
\end{equation}
For small charge and tension, we have
$$
\sigma_1 \approx \sigma - {e^2\over 4 m},\quad\quad (e^2  \ll 2 m
\sigma).
$$
Due to the scalar field dressing, the effective tension of the
brane, denoted by $\sigma_1$, is smaller than the parameter
$\sigma$ which appears in the action. This effect becomes more
dramatic if we place a large number $k$ of branes on top of each
other. Since both $\sigma$ and $e$ scale like $k$, the effective
tension of the stack is given by
\begin{equation}
\sigma_k \approx k\left[\sigma - k {e^2\over 4
m}\right],\quad\quad (k\ e^2  \ll 2 m \sigma), \label{sigmak}
\end{equation}
which grows with $k$ but less than linearly. In the limit of very
large $k$, using (\ref{transcendental}) with $\sigma$ replaced by
$k\sigma$ we have
\begin{equation}
\sigma_k \approx {m \sigma^2 \over e^2} \ln^2 \left({k\ e^2 \over
2 m \sigma}\right), \quad\quad ( k\ e^2 \gg 2 m \sigma)
\label{logarithm}
\end{equation}
and so the tension almost saturates, growing only logarithmically
with $k$. This last expression should not be taken too literally,
however, since in (\ref{actiontoy3}) we are neglecting gravitational
effects. As shown above, nonlinear effects of gravity become important at a distance given by $M_p^{2}/\alpha^2 \sigma_k$, which in the limit given by (\ref{logarithm}) is smaller than $m^{-1}$. Hence the cusp solution (\ref{cusp}), which has typical width $\sim m^{-1}$, will receive sizable gravitational corrections in the limit of large $k$.
Nevertheless, it still seems likely that the effective tension of the
stack of branes will grow with $k$ much slower than linearly.
Physically, the reason is that the scalar charge of the stack
increases with $k$. According to (\ref{junctionphi}), this means that
the cusp in the field $\varphi$ on the branes grows stronger, which
in turn means that the value of the field $\varphi_0$ on the branes
has to be further displaced into negative values. Hence, the brane
contribution to the effective tension $k\sigma e^{\alpha\phi_0}$
shows only a very modest growth with $k$, and the tension
for large $k$ is in fact dominated by the potential and gradient
energy of the scalar near the brane.

Let us now look at the interaction potential between two branes
separated from each other. For simplicity, we shall
restrict attention to the case of small scalar charge, $e^2 \ll 2
m \sigma$. Then, the last term in (\ref{actiontoy3}) is well
approximated by its linearized expression:
\begin{equation}
S= -{M_p^2\over 2}\int d^4 x \left[
(\partial \varphi)^2 + m^2 \varphi^2 \right]
- \int_{\Sigma} d^3 \xi \sqrt{\gamma}\ (\sigma + e M_p\ \varphi).
\label{actiontoy}
\end{equation}
Placing the two branes at $z=\pm d/2$, the solution for the scalar
field has the ``Golden Gate" profile shown in Fig. 3
\begin{equation}
\varphi= -{e\over m M_p}\ e^{-md/2} \cosh mz, \quad \quad  |z|<
d/2 \label{golden}
\end{equation}
$$
\varphi= -{e\over m M_p} \cosh (md/2)\ e^{-m|z|}. \quad \quad
|z|> d/2
$$
The energy per unit area of this configuration, as a function of
the interbrane distance, is given by
\begin{equation}
\sigma_2(d) = 2 \sigma_1 - {e^2\over 2 m} e^{-md}.
\label{interaction1}
\end{equation}
From this expression, and adding the long range contributions from
gravity and the four-form, the interaction potential per unit
surface is given by
$$
{\cal V}(d) = -\left[{q^2 \over 2} + {3\sigma^2 \over 4 M_p^2}
\right]\ d - {e^2\over 2 m} e^{-md}.
$$
At short distances, this takes the form
\begin{equation}
{\cal V}(d) = -{e^2\over 2m} + Q^2\ d - {m e^2 \over 4}\ d^2 + ...
\label{interaction2}
\end{equation}
where the parameter
\begin{equation}
Q^2\equiv \left[{e^2 - q^2\over 2} - {3 \sigma^2 \over 4
M_p^2}\right],\label{Q22}
\end{equation}
was introduced in (\ref{Q2}). As shown in Section III, dimensional
reduction from the supergravity lagrangian with BPS charges gives $Q^2=0$.
In this case, the linear term in (\ref{interaction2}) disappears.
The quadratic term is negative, which means that the stack of coincident
branes is unstable and tends to dissolve. Thus the stabilization of the
modulus $\varphi$ seems to make the
superposition of branes an unstable configuration.

As we shall see in more detail in Section VII, stacks of branes in
marginal or unstable equilibrium will not be appropriate for
constructing instantons, since in particular these would have too
many negative modes. The instanton is only meaningful if the
branes attract each other. The above analysis shows that at the
classical level, the branes with BPS values of the charges would
not attract each other, and consequently the nucleation of
coincident branes is not allowed at least in the semiclassical
description.

There may be several escape routes to this
conclusion. For instance, after supersymmetry breaking, the
charges of the branes get renormalized, and it is possible that
the corrected charges satisfy $Q^2>0$. In this case the branes in
the stack would attract each other with a linear potential. Other
mechanisms by which nearby branes attract each other are
conceivable, but here we shall not try to pursue their study. It
should be emphasized, however, that this remains an important open
question which needs to be addressed in order to justify the
semiclassical description of coincident brane nucleation. In
section VII, where we discuss the degeneracy factor, we shall
simply postulate that an attractive interaction exists at short
distances.

\begin{figure}[tbh]
\psfrag{f}{$\varphi$} \psfrag{d}{$d$} \centering \epsfysize=5cm
\leavevmode \epsfbox{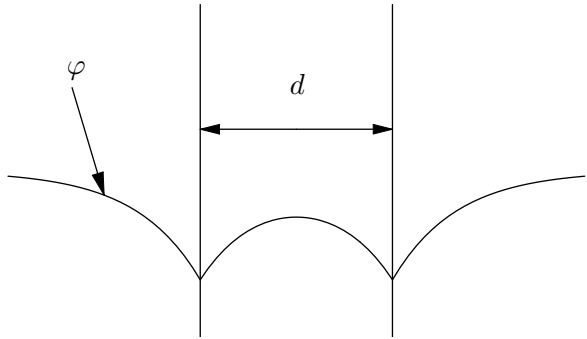}
 \caption{Profile of a massive dilaton in the presence of two branes.} \label{goldengate}
\end{figure}

\section{Coleman-de Luccia (CdL) instantons}

The brane nucleation rate per unit volume is given by an
expression of the form \cite{coleman}
\begin{equation}
\Gamma = D e^{-B}, \label{nucrate}
\end{equation}
where $B=S_E(I)-S_E(B)$. Here $S_E(I)$ is the Euclidean action of
the instanton corresponding to the decay of the four-form field,
and $S_E(B)$ is the action for the background solution before
nucleation. The prefactor $D$ will be discussed in the next
section. The formal expression (\ref{nucrate}) can be used both in flat space and in curved space. Also, it
can be used at zero or at finite temperature. The
difference is in the type of instanton and background solutions
to be used in each case.

In this section, we shall concentrate
in the maximally symmetric instanton. In flat space, this represents the decay of a metastable vacuum at zero temperature. The Euclidean solution can be described as follows
\cite{coleman}. At infinity, the field strength takes
the value $F_{o}$, which plays the role of a false vacuum. Near
the origin, we have $F=F_{i}=F_{o}-q$. This plays the role of a
true vacuum phase. Both phases are separated by the Euclidean
worldsheet of the brane, which is a three-sphere of radius
\begin{equation}
R = {3 \sigma\over \epsilon}. \label{R}
\end{equation}
Here, $\sigma$ is the tension of the brane and
\begin{equation}
\epsilon = {1\over 2}(F^2_o-F^2_i)= q [F_o -(q/2)]
\end{equation}
is the jump in the energy density accross the brane. The
difference in Euclidean actions between the instanton and the
background solution is given in this case by \cite{coleman}
\begin{equation}
B^{(flat)}\approx{27\pi^2\over{2}}{\sigma^4\over{\epsilon^3}}.
\label{Bflat}
\end{equation}
If instead of considering a single brane, we are looking at the
nucleation of $k$ coincident branes, an analogous solution of the
Euclidean equations of motion should be considered. The only
difference is that the effective charge is a factor of $k$ higher,
and the effective tension is also higher by approximately the same
factor. Using (\ref{sigmak}), we should replace
$$\epsilon \to \epsilon_k = kq [F_o -(kq/2)],$$
\begin{equation}
\sigma \to \sigma_k \approx k [\sigma-(ke^2/4m)], \label{sk2}
\end{equation}
in Eq. (\ref{Bflat})
for the ``bounce" action. As shown in Section V, the
approximate form of $\sigma_k$ is valid when the second term in
the r.h.s. of (\ref{sk2}) is small compared with the first. In the
model considered in Section IV, the modulus is stabilized with a
mass of order $m\sim M_{11}$. This may be regarded as a
conservative upper bound for the general case. Since $e=\alpha\sigma/M_p$, the condition $(k e^2/4m) \ll
\sigma$ requires
\begin{equation}
k\ll {1\over \alpha^2} \left({M_{11}^3\over
\sigma}\right)\left({M_p\over M_{11}}\right)^2. \label{kll}
\end{equation}
For larger $k$, gravitational corrections to the brane profile
become important on a lengthscale comparable to the inverse mass
of the dilaton, as we discussed in the previous section. In this
case the instanton solution will depend on the detailed dynamics
of the dilaton near the brane. The investigation of these
dilatonic instantons is per se an interesting problem, which we
leave for further research. Here, for simplicity, we shall assume
a scenario where either $\sigma \ll M^3_{11}$, due perhaps to the
wrapping of branes on a degenerating cycle, or where the extra
dimensions are relatively large, so that $M_p\gg M_{11}$, and we
shall restrict attention to instantons where the number of branes
is bounded by (\ref{kll}). This gives
\begin{equation}
B^{(flat)}_k \approx {27\pi^2\over{2}}{k\ \sigma^4\over
q^3[F_o-(kq/2)]^3}. \label{Bflatk}
\end{equation}
Nucleation in flat space is impossible when the number of branes
is too large, since otherwise we would be jumping to vacua with a
higher energy density. Therefore, we must restrict to $k < 2
F_o/q$. In fact, the minimum value of $F_i$ is achieved for the
largest $k$ satisfying $k < (F_o/q) + (1/2)$. Note that the action
increases faster than linearly as we increase the number of branes
$k$, but we still have
$$
B^{(flat)}_k \sim k \ B^{(flat)}
$$
throughout this range.

\begin{figure}[tbh]
\psfrag{Ho}[][r]{$H_o$} \psfrag{Hi}[][l]{$H_i$}
\psfrag{brane}[][]{${\rm brane}$} \centering \epsfysize=5cm
\leavevmode \epsfbox{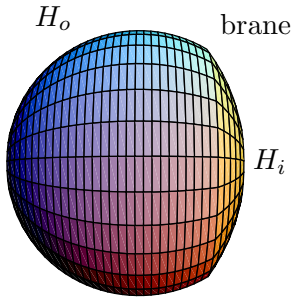}
 \caption{Coleman-de Luccia instanton, which is obtained by gluing
 together two different four-spheres of radii $H_o^{-1}$ and $H_i^{-1}$ at
 the worldsheet of the brane, itself a three-sphere.} \label{coldel}
\end{figure}

When gravity is taken into account, the maximally symmetric
instanton was given by Coleman and De Luccia \cite{deLuccia}. It is constructed by
gluing two different de Sitter solutions (that is, two
four-spheres) at the worldsheet of the brane, which is still a
three-sphere. The instanton is sketched in Fig. \ref{coldel}. The four-spheres have the radii $H_o^{-1}$ and
$H_i^{-1}$, where
\begin{equation}
H_o^2={\Lambda_{eff} \over 3 M_{P}^2},\quad\quad  H_i^2
={\Lambda_{eff} -\epsilon \over 3 M_{P}^2}, \label{Hn}
\end{equation}
are the Hubble rates of the de Sitter phases before and after the
nucleation event, respectively. Here, and in what follows, we are
assuming that the effective cosmological constant is still
positive after nucleation, since these are the final states we are
interested in. The bubble radius at the time of nucleation (which
coincides with the radius of the three-sphere) is bounded by
$0<R<H_o^{-1}$. Analytic expressions for $R$ are given in
Refs.\cite{deLuccia,teitelboim}. The general expression is
cumbersome and not particularly illuminating, so we shall
concentrate on a few limiting cases of interest.

As discussed in Section III, the gravitational field of a brane is
repulsive, and is characterized by an ``acceleration" of order
$\sigma/M_P^2$. This gravitational field will be negligible
provided that the corresponding Rindler radius (or inverse of the
acceleration) is much larger than the radius $R$ of the Euclidean
worldsheet, which in turn is smaller than $H_o^{-1}$
\begin{equation} \sigma\ll M_P^2 H_o. \label{branegravity} \end{equation}
In this regime, we can distinguish two cases. For $\sigma
H_o/\epsilon\ll 1$, the radius of the Euclidean worldsheet is much
smaller than the de Sitter radius, and the flat space expression
(\ref{Bflat}) holds. In the opposite limit, $\sigma
H_o/\epsilon\gg 1$, we have $(H_o^{-1}-R)\ll H_o^{-1}$ and
\begin{equation} B^{(wall)}\approx 2\pi^2\sigma H_o^{-3}.
\label{Bwall}
\end{equation}
The vacuum energy difference $\epsilon$ is unimportant in this
case, and the action coincides with that for domain wall
nucleation \cite{basu}.

Finally, the gravitational field of the brane is important when
\begin{equation} \sigma\gg M_P^2 H_o. \label{branegravity2}
\end{equation}
In this case, the radius of the worldsheet is given by $R\approx
(1/2\pi G\sigma)$, and
\begin{equation}
B^{(wall)}\approx {\pi \over GH_o^2} . \label{Bwallgrav}
\end{equation}
In this limit, the action of the instanton is much smaller than
the action of the background, and this is the reason why
(\ref{Bwallgrav}) is independent of the tension.

The same arguments apply of course to the instantons with coincident branes, and the corresponding expressions for the action and radii in the different regimes are summarized in Table I.

\begin{table}
\begin{tabular}{|l|l|l|}
\hline & $\sigma_k\ll M_p^2H_0$ & $\sigma_k\gg M_p^2H_0$ \\ \hline
$\sigma _{k}H_{0}/\epsilon _{k}\ll 1$ &
\begin{tabular}{l}
$R\approx 3\sigma _{k}/\epsilon _{k}$ \\
$B\approx B_{k}^{(\mbox{flat})}$%
\end{tabular}
& $R\approx 1/2\pi G\sigma_k $ \\ \cline{1-2} $\sigma
_{k}H_{0}/\epsilon _{k}\gg 1$ &
\begin{tabular}{l}
$R\approx H_{0}^{-1}$ \\
$B\approx 2\pi ^{2}\sigma_k H_{0}^{-3}$%
\end{tabular}
& $B\approx \pi /GH_{0}^{2}$ \\ \hline
\end{tabular}
\caption{Values of the radius $R$ of the Coleman-de Luccia
instanton and the corresponding bounce action $B$ in different
limits}
\end{table}

\section{The prefactor for CdL instantons}

The prefactor $D$ in Eq. (\ref{nucrate}) is given by
\cite{coleman}
\begin{equation}
D= {Z'\over Z_B}. \label{pref}
\end{equation}
Here, $Z$ and $Z_B$ are the Gaussian integrals of small
fluctuations around the instanton and the background solutions
respectively. Expanding all brane and bulk fields (which we
generically denote by $\phi$) around the instanton configuration
$\phi_I$ as $\phi=\phi_I+\sum_j\delta \phi_j$, we have $
S_E[\phi]= S_E[I] + S^{(2)}[I,\delta\phi_j] + ... , $ where
$S^{(2)}$ includes the terms quadratic in $\delta\phi_j$. At the
one loop order we have
$$
Z' = \int \Pi_j{\cal D}'(\delta\phi_j)\ e^{-S^{(2)}}.
$$
In the functional integral, there are some directions which
correspond to spacetime translations of the instanton. The primes
in the numerator of (\ref{pref}) and in the previous equation
indicate that the translational zero modes are excluded from the
integration, and replaced by the corresponding spacetime volume.
The latter is subsequently factored out in order to obtain a nucleation rate per unit time and volume.

The degrees of freedom which live on the brane will make a
contribution to the numerator but not to the denominator.
Consider, for instance, a free bosonic field $\Phi$ of mass $m_\Phi$
living on the worldsheet. Its contribution to the prefactor is
$$
D_{\Phi}=Z_{\Phi}= e^{-W_{\Phi}} = \int {\cal D}\Phi\ e^{\ \int
\gamma^{1/2} \Phi\ (\Delta^2 - m_\Phi^2)\ \Phi d^3 \xi}
$$
Here, the integral in the exponent is over the Euclidean
worldsheet of the brane. If we have $k^2$ of such fields, their
effect on the nucleation rate is to replace
\begin{equation}
B_k \to B_k + k^2 W_{\Phi},  \label{rep}
\end{equation}
in the naive expression for the nucleation rate, $\Gamma\sim
e^{-B_k}$, a replacement which can become very important as we
increase the number of fields. This is, in essence, the
observation made by Feng et al. that the large number of
worldsheet fields might strongly affect the nucleation rate.

\subsection{Scalars}

The Euclidean worldsheet in the Coleman-de Luccia instanton is a
3-sphere. Determinantal prefactors due to scalar fields were
considered in some detail in \cite{nucrates}. They are given by
\begin{equation}
W_\Phi= - \zeta'_R(-2) + (y^2/2) \ln(\sin \pi y)-{1\over
\pi^2}\int_0^{\pi y} x \ln(\sin x) dx, \label{zetadegen}
\end{equation}
where $y^2= 1- m_\Phi^2 R^2$ and $\zeta_R$ is the usual Riemann
Zeta function. For instance, the contribution of a conformally coupled
scalar field can be obtained by taking $m_\Phi^2=(3/4) R^{-2}$,
which gives
$$
W_c= - \zeta'_R(-2) + {1\over 8} \ln 2 - {7\over
16\pi^2}\zeta_R(3) \approx 0.0638
$$
Hence, the effective degeneracy factor contributed by a conformal
scalar field is given by
\begin{equation}
D_{c}\approx e^{-W_c} \approx 0.94 < 1. \label{sup}
\end{equation}
The first thing to note is that this factor is not an enhancement,
but a suppression. Hence, the prefactor cannot simply be thought
of as the exponential of an entropy. More generally, from Eq.
(\ref{zetadegen}), the prefactor is independent of the expansion
rate of the ambient de Sitter space. This has implications for the
scenario proposed by FMSW, as we shall discuss in the next
Section.

\begin{figure}[tbh]
\psfrag{Ds}[][]{$D_s$}\psfrag{mR}[][]{$^{}\ \ \ m_\Phi R$}
 \centering
\epsfysize=5cm \leavevmode
\epsfbox{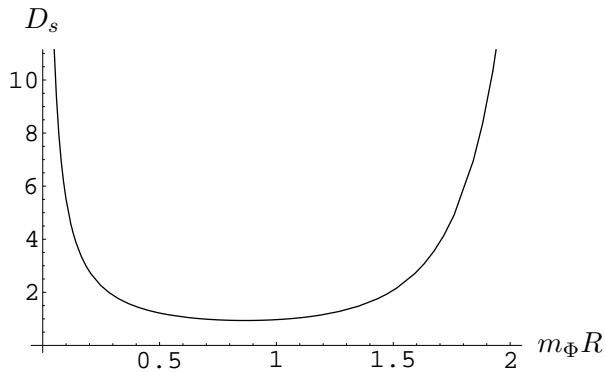} \caption{Contribution of a scalar field to the prefactor $D$ in the nucleation rate (\ref{nucrate}), as a function of its mass $m_\Phi$ (measured in units of the inverse radius of the instanton.) At low mass, the enhancement is due to large phase space:  bubbles can nucleate with values of the field in the range (\ref{delphik}), which becomes larger for smaller masses. The enhancement at large mass can be understood as a finite renormalization of the brane tension.} \label{prefactor}
\end{figure}

In general, the degeneracy factor will depend on $R$ and on the
mass of the particle. For light minimally coupled scalars,
equation (\ref{zetadegen}) gives
\begin{equation}
D_s \approx {e^{{\zeta_R}'(-2)}\over \pi^{1/2} m_\Phi R} \quad
(m_\Phi R\ll 1). \label{inv}
\end{equation}
There can be a strong enhancement in the nucleation rate if there
are very light scalar fields. In the limit $m_\Phi \to 0$
the factor goes to infinity. This is because a massless scalar has
a normalizable zero mode on the sphere, corresponding to the
symmetry  $\Phi \to \Phi + const$. In this case, the zero mode
must be treated as a collective coordinate. The nucleation rate is
proportional to the range $\delta\Phi$ of the field $\Phi$,
because the bubbles can nucleate with any average value of the
scalar field with equal probability \cite{nucrates}
\begin{equation}
D_s(m_\Phi^2=0) = \lim_{m_\Phi^2\to 0} [m_\Phi D_s(m_\Phi)] (\pi
R^3)^{1/2} \delta \Phi = e^{{\zeta_R}'(-2)} R^{1/2} \delta \Phi.
\label{dsmassless}
\end{equation}
Finally, for large mass the expression (\ref{zetadegen}) leads to
\begin{equation}
D_s \approx \exp({\pi m^3_\Phi R^3/6}) \quad \quad (m_\Phi R \gg
1). \label{renorm}
\end{equation}
The exponent of this expression can be interpreted as a renormalization of the tension of the stack of branes, due to the heavy scalars living on it \cite{gm}. Indeed, the effective potential for a scalar field in 2+1 dimensions in the flat space limit is proportional to $m_\Phi^3$, and the factor of $R^3$ is just due to the volume of the worldsheet.
The factor $D_s$ is plotted in Fig. \ref{prefactor} for different values of $m_\Phi R$.

Note that there is an enhancement both at smal and at large
mass, but the two have very different origin. The large value of (\ref{inv}) for light fields can be interpreted as a phase space enhancement. As we shall discuss in Subsection VII-D, quantum fluctuations of fields living
on the worldsheet of the brane are characterized by a temperature
$T= 1/2\pi R$. The corresponding fluctuations in the potential term are
of order $m_\Phi^2 \Phi^2 \sim T^3$, which corresponds to a root mean squared expectation value for $\Phi$ of the order
\begin{equation}
\delta\Phi \sim \frac {1}{m R^{3/2}}. \label{delphik}
\end{equation}
Eq. (\ref{inv}) is recovered if we insert the range (\ref{delphik}) in Eq. (\ref{dsmassless}). The lighter the field, the larger is the phase space factor $\delta\Phi$, and the larger is the nucleation rate. For $m_\Phi R \gg 1$ this argument cannot be used,
since the field does not behave as effectively massless. In that limit the field decouples, as it should, and its effect is felt as a renormalization of the parameters in the lagrangian. For a scalar field, the leading effect is to renormalize the tension of the brane, making it lower (See Appendix A). This causes an exponential enhancement of the nucleation rate.

\subsection{$U(k)$ fields}

Besides scalar fields, gauge bosons and fermions may live on a
brane. At low energies, the field content on a stack of coincident
branes will be model dependent. The idea is that it will correspond to a gauge theory whose symmetry is enhanced when
branes are coincident, giving rise to a large number of light
species on the worldsheet. The details of the theory, however, will depend on whether we start from 2D
branes which descend directly from the ten dimensions, of from
higher dimensional p-branes wraped on (p-2)-cycles. They will also
depend on the details of compactification. Rather than building a
particular scenario from first principles, here we shall try to
gain some intuition by considering a toy model directly in four
dimensions. The degrees of freedom on the stack of $k$ coindident
branes will be bluntly modeled by a weakly coupled SUSY $U(k)$ gauge
theory on the 2+1 dimensional worldsheet. This contains a $U(k)$
gauge field, $(k^2-1)$ scalar degrees of freedom in the adjoint
representation of $SU(k)$, and a scalar singlet; plus the
corresponding fermionic degrees of freedom. The action for the worldsheet fields is given by
\begin{equation}
S_{SYM} = -\int \sqrt{\gamma} d^3\xi \left[{1\over 2}{\rm Tr}(F_{\mu\nu}F^{\mu\nu}) + {\rm Tr}(D_\mu\phi D^\mu\phi) + {\cal V(\phi)} + ...\right],
\label{ym}
\end{equation}
where the ellipsis indicate the terms containing fermions.
Here $F_{\mu\nu}$ is the field strength of the gauge field $A^\mu=A^\mu_a \lambda_a$, where $\lambda_a$ are the generators of $U(k)$, normalized by ${\rm Tr}(\lambda_a\lambda_b)=\delta_{ab}/2$, and $\phi=\phi_a\lambda_a$, where $\phi_a$ are real scalar fields $(a,b=1,...,k^2)$.
By analogy with the well known case of D-branes in 10 dimensions, we shall assume that if the coincident branes are flat (as in the case when there is no external four-form field), then the theory is supersymmetric and all degrees of freedom are massless. The scalar field $\phi$ is a hermitian matrix and can always be diagonalized by a suitable $U(k)$ gauge transformation, $\phi={\rm diag}(\varphi_1,...,\varphi_k)$. The eigenvalues $\varphi_i$, $i=1,...,k$ are then interpreted as the positions of the different parallel branes along an axis perpendicular to them (if the codimension of the branes were higher, there would be additional scalar matrices representing the positions of the branes along the additional orthogonal directions, but here we are interested in the case of codimension one). In the supersymmetric case, the potential for the scalar field vanishes,
$
{\cal V}(\phi)=0.
$
However, for non-flat branes, the displacement of the stack of branes no longer behaves as an exatly massless field \cite{nucrates,covariant}, but one which couples to a combination of the worldsheet and extrinsic curvatures, as well as to the background four-form field. Also, after moduli stabilization, the forces amongst branes are nonvanishing, and this also contributes to the potential for the relative displacements of the different branes. Thus, as we shall see, there will be a nonvanishing potential ${\cal V}(\phi)\neq 0$ in the physical situation of our interest.

When the positions are not coincident, the $U(k)$ symmetry breaks to a smaller group because some of the gauge bosons aquire masses $m^2_A=(g^2/2)(\varphi_i-\varphi_j)^2$. There are always at least $k$ massless vectors (corresponding to Maxwell theory on the individual branes on the stack) and the remaining $k^2-k$ have double degeneracy. For example, in the
case of a single brane, the scalar field will
represent the Goldstone mode of the broken translational symmetry,
associated to transverse displacement of the brane. For the case
of two branes, there are two such scalars. One of them, $\varphi_+=(\varphi_1+\varphi_2)/\sqrt{2}$, will
corresponds to simultaneous motion of both branes, and is a
singlet under $SU(2)$. The other one, $\varphi_-=(\varphi_1-\varphi_2)/\sqrt{2}$, will correspond to relative
motion of the branes and it transforms under $SU(2)$. When the two branes
move apart, $\phi_-$ acquires an expectation value and
two of the gauge bosons get a mass, breaking the symmetry $U(2)
\to U(1)\times U(1)$.

The case of interest to us is not a flat brane, but the
world-wolume of the $2$-brane in the CdL instanton, which forms a
3-sphere of radius $R$. In this situation, we do not expect the
theory to be supersymmetric (in particular, corrections to the effective action will appear at one loop, which will be related in fact to the determinantal prefactor in the nucleation rate).
The case of a single brane is very similar to the case of a vacuum
bubble, and in that case we know that the transverse displacements
correspond to a scalar field of negative mass squared \cite{covariant,nucrates}
\begin{equation}
m^2_+=-3 R^{-2}. \label{nmass}
\end{equation}
The origin and precise value of this mass term can be understood
geometrically, since it leads to four normalizable zero modes
which are the spherical harmonics with $l=1$. These correspond to
the four space-time translational modes of the instanton, which
have to be treated as collective coordinates.
This scalar field has also a single negative mode, which
is the constant $l=0$ mode. A negative mode is precisely what is
needed for an instanton to contribute to the imaginary part of the
vacuum energy, and hence to contribute to false vacuum decay
\cite{coleman}.
Integrating out the transverse displacement of the brane gives a
determinantal prefactor of the form \cite{nucrates}
\begin{equation}
D_+= {\sigma^2 R^2 \over 4} e^{\zeta'_R(-2)} \Omega,
\label{dgoldst}
\end{equation}
where $\Omega=VT$ is the spacetime volume.  The prefactor in
the nucleation rate (\ref{nucrate}) per unit time and volume is obtained after dividing by $\Omega$. The above argument neglects gravity, and it is a
good approximation when $\sigma_0 \ll M_p^2 H_o$. The case of
strong gravity $\sigma_0 \gg M_p^2 H_o$ is far more complicated,
since one has to integrate over fluctuations of the gravitational
field in the bulk, and is left for further research.

If there are 2 coincident branes, then there are two independent transverse displacements corresponding to the eigenvalues
$\varphi_1$ and $\varphi_2$. The ``center of mass" displacement $\varphi_+$ behaves just like in the case of a single brane. The orthogonal combination $\varphi_-$ represents the brane separation, and the two-brane instanton will only be relevant if this second combination acquires a positive mass through some mechanism, so that there is a single negative mode, not two, and four normalizable zero modes in total. In other words, the branes must attract each other.
As shown in Section V, if the branes have BPS charges, they in
fact tend to repel each other once the dilaton is stabilized, and
the configuration with coincident branes is unstable. In this
case, we do not expect that there will be any instanton
representing the nucleation of multiple branes.

One possible way out of this conclusion, is to assume that the charges are different from their BPS values, due to supersymmetry breaking effects, so that the sum $Q^2$ defined in Eq. (\ref{Q22}) is
positive. In that case, the two branes attract each other with a
linear potential. The canonical field $\varphi_-$ is related to the
interbrane distance $d$ by $|\varphi_-| = \sigma^{1/2} d$. Hence, the
interbrane potential (\ref{interaction2}) takes the form
\begin{equation}
{\cal V}(\varphi_-) = Q^2 \sigma^{-1/2}|\varphi_-| - {m e^2\over 4 \sigma}
\varphi_-^2 + ...\label{interaction3}
\end{equation}
This potential is attractive at small distances, and has a maximum
at $\varphi_-= \varphi_m \sim Q^2 \sigma^{1/2}/m e^2$. Classically, the
branes will attract at short distances. However, there is a danger
that they will be separated by quantum fluctuations. As we discussed in the previous Subsection (and we will argue more at length in Subsection D) quantum fluctuations of fields living
on the worldsheet of the brane are characterized by a temperature
$T= 1/2\pi R$. The fluctuations in the potential are
of order ${\cal V} \sim T^3$, and these correspond to a root mean
squared expectation value for $\varphi_-$ of the order
\begin{equation}
\delta\varphi_- \sim \sigma^{1/2}/R^3 Q^2. \label{deltaphim}
\end{equation}
The stability of the two-brane instanton requires that
$\delta\varphi_- \ll \varphi_m$. Otherwise, unsuppressed quantum
fluctuations take the field over the barrier and the distance
between the branes starts growing without bound. This requires
\begin{equation}
Q^4 \gg {m e^2 \over R^3}. \label{largeQ}
\end{equation}
If this condition is satisfied, then $\varphi_-$
is trapped near the origin, and the branes stay together. It can be shown that for $Q^4 \gg \sigma R^{-5}$ the field
behaves as approximately
massless in the range given by (\ref{deltaphim}), so from
(\ref{dsmassless}) its contribution to the prefactor can be
estimated as
\begin{equation}
D_- \sim R^{1/2} \delta\varphi_- \sim {\sigma^{1/2} \over R^{5/2}
Q^2}.
\label{dml}
\end{equation}
As in the case of the massive field discussed in the previous Section, this expression is only justified when $D_- \gg 1$. If $Q^2$ is too large, then the field will not behave as massless in the range (\ref{deltaphim}), and we expect that
for $Q^4 \gg \sigma R^{-5}$ the sole effect of the field will be to renormalize the coefficients of operators such as the brane tension in the classical lagrangian. An inconvenient feature of the linear potential (\ref{interaction3}) is that
is non-analytic at the origin, and hence an explicit calculation in the limit of large slope is not straightforward. Moreover, we cannot write down an expression for it in terms of the matrix operator $\phi$, but just in terms of its eigenvalues $\varphi_i$.

Another possibility, which is more tractable from the formal point of view,
is to assume that there is an attractive interbrane potential which is
quadratic at short distances. That is, as in (\ref{interaction3}) with
$Q^2=0$ but with a positive coefficient in front of the second term.
In terms of the eigenvalues $\varphi_i$, which represent the displacements of
the branes, we assume the following expression for the potential
\begin{equation}
{\cal V}(\varphi_j) = m_+^2 \varphi_+^2 +
{1\over 2k} m_-^2 \sum_{ij} (\varphi_i-\varphi_j)^2 + ...,
\end{equation}
where $\varphi_+ = k^{-1/2}\sum_{i=1}^k \varphi_i = k^{-1/2}{\rm Tr}(\phi)$
and $m_+^2=-3R^{-2}$ [as given in Eq. (\ref{nmass})], while $m_-^2>0$ is a new parameter
which characterizes the attractive interaction at short distances.
In terms of the field $\phi$, we can write the potential as
\begin{equation}
{\cal V}(\phi) = m_+^2 k^{-1}({\rm Tr}\phi)^2 +
m_-^2 [{\rm Tr}\phi^2-k^{-1} ({\rm Tr}\phi)^2]
\end{equation}
$$
= {1\over 2} m_+^2 \phi_1^2 +
{1\over 2}m_-^2 \sum_{b=2}^{k^2} \phi_b^2,
$$
where in the last equality we have expanded $\phi=\phi_a\lambda_a$ in the basis
of generators $\lambda_a$, and we have used $\lambda_1 = (2k)^{-1/2}{\bf 1}$ and
${\rm Tr}\lambda_b = 0$ for $b=2,...,k^2$. In the symmetric phase (and assuming $m_-R\ll 1$)
each one of the adjoint fields $\phi_b$ will contribute a determinantal prefactor of the form
\begin{equation}
D_- \approx {e^{{\zeta_R}'(-2)}\over \pi^{1/2} m_- R}, \quad (m_-R\ll 1) \label{dmq}
\end{equation}
where we have used (\ref{inv}). This shows a somewhat milder dependence in $R$ than in the case of a linear interaction between branes, given in Eq. (\ref{dml}), but still of power law form. In the limit of large mass $
D_- \approx \exp({\pi m^3_- R^3/6})$, which as discussed before amounts to a finite renormalization of the brane tension.

Aside from scalars, we should also consider the contributions from
gauge bosons and fermions. For the case of a 3-sphere, these have
been studied in \cite{gm}. For vectors of mass $m_A$, the result
is
\begin{equation}
W_{A} = -{1\over 2}\log\left( {g^2 R\over
4\pi^2}\right)-\log\left({\sinh(\pi m_A R)\over \pi m_A
R}\right)-\int_0^{m_A R} y^2 {d\over dy} \log(\sinh \pi y) dy -
   \zeta'_R(-2)+2\zeta'_R(0),
\end{equation}
where $g$ is the gauge coupling, which is dimensionful in three
dimensions. When the branes are coincident, the theory is in
the symmetric phase and the gauge bosons are massless. A massless
gauge boson gives a contribution of the form
\begin{equation}
D_A= e^{-W_A}= g R^{1/2}
e^{\zeta'_R(-2)},   \quad\quad (m_A=0)
\label{da}
\end{equation}
which again behaves as a power of $R$. A Dirac fermion of mass
$m_\Psi$ yields the contribution \cite{gm}
\begin{equation}
W_\Psi = {1\over 4} \log\cosh(\pi m R) + \pi \int_0^{mR} u^2
\tanh(\pi u) du + 2\zeta'_R(-2,1/2)-{1\over 2} \zeta'_R(0,1/2).
\label{wpsi}
\end{equation}
For the massless case, the $R$ dependent terms vanish and we have
\begin{equation}
D_\Psi = 2^{-1/4} e^{{3\over 2} \zeta_R'(-2)}, \quad\quad\quad (m_\Psi =0).
\label{dpsi}
\end{equation}
which is a constant independent of the radius $R$.

\subsection{Nucleation rate}

Collecting all one-loop contributions, the prefactor in Eq.(\ref{nucrate})
due to the weakly coupled $U(k)$ gauge theory in the unbroken phase is given by
\begin{equation}
D={D_+\over \Omega} (D_-)^{k^2-1} (D_A)^{k^2} (D_\Psi)^{k^2}.
\end{equation}
Using (\ref{dgoldst}), (\ref{dmq}),(\ref{da}) and (\ref{dpsi}) we
are led to a nucleation rate per unit volume of the form
\begin{equation}
\Gamma_k \approx {\pi^{1/2}m_-\sigma_k^2 \over 4} \ R^{3}
\left({e^{7\zeta'_R(-2)}g^2\over \sqrt{2} \pi m_-^2
R}\right)^{k^2/2} e^{-B_k},  \quad (m_- R\ll 1) \label{finald}
\end{equation}
where $B_k$ is the corresponding bounce action for the nucleation
of $k$ coincident branes. Note that the $R$ dependence of the
prefactor is simply as a power law. Here we have used the form
(\ref{dmq}) for the scalar contribution $D_-$, corresponding to an
attractive interaction amongst branes which is quadratic at short
distances, with a curvature of the potential characterized by some
mass parameter $m_-$ \footnote{If we assume instead a linear interaction at
short distances, we should use (\ref{dml}) and the behaviour
changes to $D\propto R^{(9/2)-2k^2}$, but in any case the dependence
is still a power of the radius $R$}.
The prefactor in Eq. (\ref{finald}) has the exponential dependence
on $k^2$ which counts the number of worldsheet field
degrees of freedom, while the Euclidean action $B_k$ behaves
approximately linearly with $k$. Hence, as suggested in
\cite{FMSW}, the prefactor can be quite important in determining
the nucleation rate.

In the scenario proposed in \cite{FMSW} it
was also desirable that the enhancement in the nucleation rate would switch off at the present time, in order to prevent the vacuum from decaying further. Unfortunately, the expression (\ref{finald}) does not seem to have this property. The prefactor depends only on the radius $R$, which is itself a function of various parameters, such as the brane tension, the charge and the ambient expansion rate, as summarized in Table I.
According to (\ref{finald}), an enhancement of the nucleation rate of coincident branes will occur for
\begin{equation}
R\sim {\rm Min}\left\{\frac {3\sigma_k}{\epsilon_k},H_o^{-1},\frac{4M_p^2}{\sigma_k}\right\} \ll (g/m_-)^2,
\end{equation}
where we have used the results of Table 1 in the first step.
Consider first the situation where $\sigma_k M_p^{-2}\ll H_o$. Note that even in the regime when $R\approx H_0^{-1}$, the dependence of the degeneracy factor on
the corresponding dS temperature is
only power law, and not exponential as suggested in \cite{FMSW}. Also, it is clear the stability of our vacuum is not guaranteed by the smallness of the present expansion rate. The enhancement will persist provided that $\sigma_k/\epsilon_k \ll (g/m_-)^2$, even if the ambient de Sitter temperature vanishes. More worrisome is the fact that for sufficiently large $k$ we enter the regime where $\sigma_k M_p^{-2}\gg H_o$. In that case, we have
$$
R \sim M_p^2/\sigma_k,
$$
which can get smaller and smaller as we increase the number of coincident branes, eventually leading to a catastrophic decay rate, regardless of the value of $H_o$.

The expression (\ref{finald}) for the nucleation rate is valid for $m_- R \ll 1$. In the opposite limit, $m_- R \gg 1$, the scalars decouple, contributing a finite renormalization of the parameters in the action (such as the brane tension and induced Newton's constant). For completeness, this is discussed in Appendix A, where it is shown that the renormalization of parameters can have a very significant impact on the nucleation rate.

\subsection{Discussion: temperature of a vacuum bubble}

In Ref. \cite{FMSW}, the prefactor in the nucleation rate for the nucleation of coincident branes was estimated as an entropy
enhancement $$ D \sim e^S, $$ where, from dimensional analysis, the entropy was estimated as $S \propto T^2 R^2$ per field
degree of freedom. The factor $R^2$ is due to the area of the
bubble, and $T$ is some effective temperature. Although the interpretation of the prefactor as the exponential of an entropy should not be taken too literally, let us try and phrase the results of the previous subsection in this intuitive language.

A particle detector following a geodesic in a de Sitter space responds
as if it was at rest in a thermal bath in flat space, at the
temperature $T_o=H_o/2\pi$. It should be kept in mind, however, that the dS invariant quantum state is in fact a pure state,
and hence rather different from a true thermal
state. For instance, any two detectors in geodesic relative motion observe the same temperature, with a perfectly isotropic distribution. This is a consequence of de Sitter invariance, and is in contrast with the situation in a thermal bath in flat space, where moving observers detect a temperature blue-shift in the direction of their motion relative to the bath.

The fields living on a nucleated brane will experience some
thermal effects too. The bubble is embedded in a dS space
characterized by a temperature $T_o$. The interior of the
bubble is also a dS space characterized by a different
expansion rate, with corresponding temperature $T_i$. The existence
of two different de Sitter spaces in contact with the brane led
the authors of Ref. \cite{FMSW} to consider two different
possibilities for the effective temperature of the fields on the
brane: $T=T_o$ and $T=(T_o T_i)^{1/2}$. However, there is in fact no ambiguity in the temperature of such fields \cite{solutions}, which is determined as follows.

The worldsheet of the brane is an $S^3$ of radius $R$, the Euclidean de Sitter
space in 2+1 dimensions. If interactions with bulk fields are
neglected, brane fields are only sensitive to the
geometry of the worldsheet, and do not know about the properties
of the ambient space. In this approximation, the relevant
temperature is clearly the intrinsic temperature of the
lower dimensional de Sitter space,
\begin{equation}
T_R= 1/2\pi R. \label{rindler}
\end{equation}
This conclusion remains unchanged when we include interactions
with bulk fields. The simplest way to see this, is to consider the
limiting case where gravity can be ignored and the nucleation takes
place in a flat space. There, the ambient temperature vanishes $T_o=0$, but
the fields on the brane will feel the temperature $T_R$, because the nucleated brane expands with constant acceleration
$a=1/R$. An accelerating observer in the Minkowski vacuum will
detect a Rindler temperature $T_R=a/2\pi$, which happens to
coincide with the intrinsic worldsheet temperature. Hence, the de
Sitter vacuum in the 2+1 dimensional worldsheet is in equilibrium
with the Minkowski vacuum in the bulk. This conclusion is quite
general, and applies also to bubbles nucleating in de Sitter. The
CdL instanton has an $O(4)$ symmetry under Euclidean rotations.
This becomes an $O(3,1)$ symmetry after analytic continuation into
Lorentzian time. The quantum state after bubble nucleation is
expected to inherit this symmetry \cite{coleman,vachalex,montes},
and the only way to achieve it is if the fields on the brane are
in their intrinsic de Sitter vacuum, which is characterized by
temperature $T_R$.

Note that $T_R$ is a relatively high temperature. The radius of
the instanton is always smaller than $H_o^{-1}$ (see Table 1), and
therefore $T_R$ is strictly larger than $T_o$ and $T_i$.
Nevertheless, the product $k^2\ T_R^2\
R^2 \sim k^2$, and hence the ``entropy enhancement", is independent of the ambient de Sitter
temperature. As shown in the previous subsections, the independence of the prefactor $D$ on the ambient expansion rate is only approximate, due to the anomalous behaviour of light fields in de Sitter space. This introduces a dependence of the effective action $W_{eff}=-\log D$ on the radius $R$ of the instanton (which in turn may depend on $H_o$ in certain regimes). This dependence, however, is quite different from the one proposed in  \cite{FMSW}, where it was suggested that the nucleation rate of coincident branes would be enhanced at high $H_o$, and would
switch off at low $H_o$ due to the drop in ambient temperature. What we find instead is that, if the CdL instanton for nucleation of coincident branes really exists, then the corresponding degeneracy factor does not necessarily switch off.

We shall return to a discussion of this point in the concluding Section. Before that, let us turn our attention to a different instanton, which may be relevant to the FMSW scenario.

\section{Pair creation of critical bubbles}

Euclidean de Sitter space is compact in all spacetime directions,
and (as we just discussed) it behaves in some respects as a system at finite
temperature. One may then ask whether there are instantons similar to the thermal ones in flat space. These correspond to static bubbles, in unstable equilibrium between expansion and collapse.

Static instantons with O(3) symmetry have previously been considered in a variety of contexts, notably for the description of false vacuum decay in the presence of a black hole (See e.g. \cite{hiscock,bkk} and references therein). The particular instanton we shall consider corresponds to pair creation of critical bubbles in de Sitter, and to our knowledge it does not seem to have received much attention in the past. This is perhaps not surprising, since its action is higher than the action for the maximally symmetric CdL instanton. However, if the CdL instanton does not exist for coincident branes, the static one may turn out to be relevant once the degeneracy factors are taken into account.

\subsection{The instanton solution}

\begin{figure}[tbh]
\psfrag{rm}[][r]{$r_+$}\psfrag{R}[][r]{$R$}\psfrag{0}[][r]{$0$}
\psfrag{t}[][r]{$t$} \psfrag{f}[][l]{$\phi$} \centering
\epsfysize=5cm \leavevmode \epsfbox{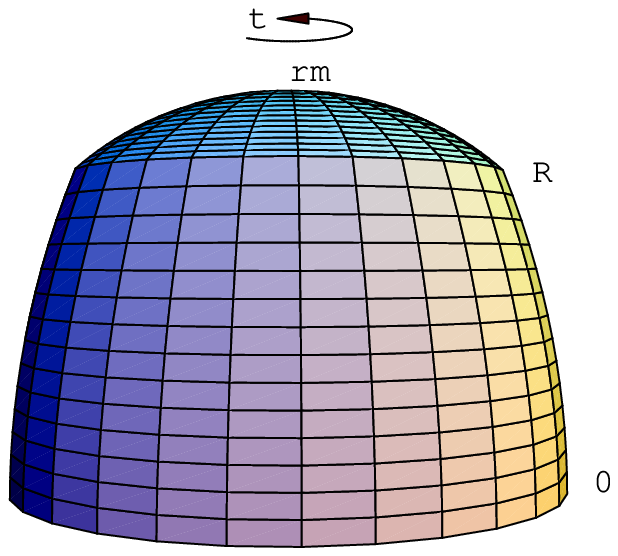}
\epsfbox{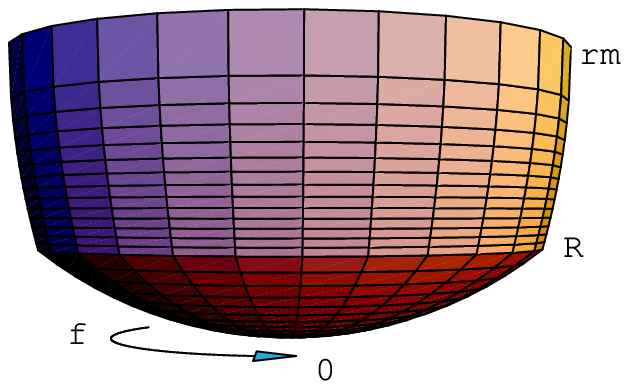} \caption{Static instanton in de
Sitter space. The left figure shows the geometry induced on the
plane $r,t$, while keeping angular coordinates fixed, whereas the
right figure shows the geometry induced on the plane $r,\phi$,
keeping $\theta$ and $t$ fixed. The vertical direction corresponds
to the coordinate $r$, common to both pictures. The cosmological
horizon is at $r=r_+$, the brane is at $r=R$, and $r=0$ is the
center of the static bubble of the new phase.} \label{estatico}
\end{figure}

The energy of a critical bubble is different from zero, and
consequently, the metric outside of the bubble is no longer pure
de Sitter, but Schwarzschild-de Sitter (SdS). The instanton is a
solution of the Euclidean equations of motion, with two metrics
glued together at the locus of the wall, which is a surface of
constant $r$ in the static chart of SdS (see Fig.~\ref{estatico}).
The metric outside is given by
\begin{equation}
ds^{2}=f_{o}\left( r\right) dt^{2}+f_{o}^{-1}\left( r\right)
dr^{2}+r^{2}d\Omega ^{2},  \label{metrico}
\end{equation}
where $d\Omega ^{2}=d\theta ^{2}+\sin ^{2}\theta d\phi ^{2}$, and
\begin{equation}
f_{o}(r)=\left( 1-\frac{2GM}{r}-H_{o}^{2}r^{2}\right) .  \label{efeo}
\end{equation}
The metric inside is given by
\begin{equation}
ds^{2}=C^{2}f_{i}\left( r\right) dt^{2}+f_{i}^{-1}\left( r\right)
dr^{2}+r^{2}d\Omega ^{2},  \label{metrici}
\end{equation}
where
\begin{equation}
f_{i}\left( r\right) =\left( 1-H_{i}^{2}r^{2}\right) ,
\label{efei}
\end{equation}
corresponding to a de Sitter solution. The constant $C$ is
determined by the condition that on the brane (i.e., at $r=R$) the
two metrics must agree, which leads to $C=[f_{o}\left( R\right)
/f_{i}\left( R\right)]^{1/2} $.

The parameters $M$ and $R$ depend on $\sigma_k $, $H_{o}$ and
$H_{i}$. Their values are determined by the junction conditions at
the brane \cite{israel},
\begin{equation}
\left[ K_{ab}\right] =-4\pi G\sigma_k \gamma _{ab},  \label{discont2}
\end{equation}
where $\left[ K_{ab}\right] $ is the difference in the extrinsic
curvature $K_{ab}=(1/2)f^{1/2}\partial _{r}g_{ab}$ on the two
sides and $\gamma _{ab}$ is the world-sheet metric.
Eq.\ (\ref{discont2}) gives rise to the junction conditions,
\begin{equation}
[g]=-4\pi G \sigma_k,\quad [g']=0, \label{juco}
\end{equation}
where we have introduced the new function $g(r)=f^{1/2}(r)/r$.
Using Eqs.~\ (\ref{efeo}) and (\ref{efei}), we have
\begin{equation}
g_{o}g_o^{\prime } =-\frac{1}{r^{3}}+\frac{3GM}{r^{4}},
\quad\quad
g_{i}g_i^{\prime } =-\frac{1}{r^{3}}. \label{cuadradosprima}
\end{equation}
Hence, using (\ref{juco}), $g_{o}^{\prime }\left( R\right)
=g_{i}^{\prime }\left( R\right)= -3M/4\pi \sigma_k R^{4}$, and then
$g_{i}(R)$ and $g_{o}(R)$ are easily obtained from
Eqs.~(\ref{cuadradosprima}):
\begin{equation}
g_{i}\left( R\right) =\frac{4\pi \sigma_k R}{3M}, \quad\quad
g_{o}\left( R\right) =g_{i}\left( R\right) \left(
1-\frac{3GM}{R}\right) . \label{ges}
\end{equation}
From (\ref{efeo}) and (\ref{efei}) we have
\begin{equation}
g_o^{2 }(R) =\frac{1}{R^{2}}-\frac{2GM}{R^{3}}-H_{o}^{2},
\quad\quad
g_i^{2 }(R) =\frac{1}{R^{2}}-H_{i}^{2}.
\label{cuadrados}
\end{equation}
Inserting (\ref{ges}) in
(\ref{cuadrados}) we finally obtain a quadratic equation for
$g_{i}\left( R\right) \equiv x$. The solution is
\begin{equation}
x=\frac{\epsilon }{4\sigma_k }+\frac{3\sigma_k }{16M_{p}^{2}}+\left[
\left( \frac{\epsilon }{4\sigma_k }+\frac{3\sigma_k
}{16M_{p}^{2}}\right) ^{2}+\frac{ H_{i}^{2}}{2}\right] ^{1/2},
\end{equation}
where we have used $H_{o}^{2}-H_{i}^{2}=8\pi G\epsilon /3=\epsilon
/3M_{p}^{2}$. Then the parameters $M$ and $R$ are given in terms
of $x$ by
\begin{equation}
R^{-2} =x^{2}+H_{i}^{2}, \quad\quad
M =4\pi \sigma_k R/3x.
\end{equation}
This concludes the construction of the instanton solution for
given values of the physical parameters $\sigma_k, H_o$ and $H_i$.

The equation $f_{o}\left( r\right) =0$ has three real solutions
for $27H_o^2M^2G^2<1$. One of them, say $r_-$, is negative and
the other two are positive. The two positive roots correspond to
the black hole and cosmological horizons. We call them
respectively $r_s$ and $r_{+}$. Therefore we can write
\begin{equation}
f_{o}\left( r\right) =-\frac{H_{o}^{2}}{r}\left( r-r_-\right)
\left( r-r_s\right) \left( r-r_{+}\right) . \label{fofactors}
\end{equation}
Of course, in our instanton the horizon at $r_s$ is not present, since the exterior metric is matched to an interior metric at some $r=R > r_s$ (see Fig. \ref{estatico}). For $r<R$ the metric is just a ball of de Sitter in the static chart, and it is regular down to the center of symmetry at $r=0$. In general, the size of the cosmological
horizon is given by
\begin{equation}
H_{o}r_{+}=\frac{2}{\sqrt{3}}\cos\left(\frac{\varphi+\pi}{3}\right),
\end{equation}
where we have introduced the angle
\begin{equation}
\varphi=- \arctan \sqrt{{1\over 27 H_o^2 M^2 G^2}-1},
\label{varphi}
\end{equation}
In the limit $M\rightarrow 0$ the angle
$\varphi\rightarrow-\pi/2$, and $H_{o}r_{+}\rightarrow 1$.

According to Eq\ (\ref{ges}), at the brane we have
$f_{o}\left( R\right) =x^{2}\left( R-3GM\right) ^{2}$, so the
equation $f_{o}\left( R\right) =0$ has a double zero instead of
two different roots. This means that the radius of the instanton
will coincide with the radius of one of the horizons only in the
special case where both horizons have the same size, $
r_s=r_{+}=R=3GM$. In the limit $ r_s=r_{+}$ the exterior metric becomes the Nariai solution \cite{nariai,perry},
which has $ r=\left( \sqrt{3}H_{o}\right) ^{-1}$. Note that in the
limit $3GM \rightarrow\left( \sqrt{3}H_{o}\right) ^{-1}$,
$\varphi\rightarrow 0$ and $H_{o}r_{+}\rightarrow 1/\sqrt{3}$, as
expected.

Like in the case of instantons describing the production of black holes \cite{perry} or monopoles \cite{basu} in de Sitter, the instanton presented here describes the creation of {\em pairs} of bubbles. As we shall see, the Euclidean solution is periodic in the time direction, so that time runs on a circle $S^1$ (See Fig. 6). The geometry at the time of nucleation is obtained by slicing the compact instanton through a smooth spacelike surface which cuts the $S^1$ factor at two places, say, $t=0$ and $t=\pi$. The resulting geometry contains two different bubbles separated by a distance comparable to the inverse expansion rate.

\subsection{Temperature and action}

In order to calculate
the temperature of the worldsheet in the static instanton we must first find the time periodicity $\beta$ . This is
determined by the regularity of the Euclidean metric at
the cosmological horizon. For $r\rightarrow r_{+}$, we have
\begin{equation}
f_o(r)\approx A^{2}\left( 1-\frac{r}{r_{+}}\right),
\end{equation}
where
\begin{equation}
A^{2}=H_{o}^{2}\left( r_{+}-r_-\right) \left( r_{+}-r_s\right)
=3H_{o}^{2}r_{+}^{2}-1 . \label{A}
\end{equation}
In terms of the new coordinates
\begin{equation}
\rho =\frac{2r_{+}}{A}\sqrt{1-\frac{r}{r_{+}}},\quad \phi
=\frac{A^{2}}{ 2r_{+}}t,
\end{equation}
the metric (\ref{metrico}) for $r\to r_+$ reads
\begin{equation}
ds^{2}=\rho ^{2}d\phi ^{2}+d\rho ^{2}+r_{+}^{2}d\Omega ^{2},
\end{equation}
so it is clear that $\phi $ is an angle, $0\leq \phi \leq 2\pi$,
and $t$ varies in the range $0\leq t\leq 4\pi r_{+}/A^{2}$.
Therefore the value of $ \beta $ is
\begin{equation}
\beta =\frac{4\pi r_+}{3 H_{o}^2 r_{+}^{2}-1}= \frac{2\pi
r_+^2}{r_{+}-3GM}.
\end{equation}
The temperature of the worldsheet instanton is given by the proper
time periodicity $ \beta _{R}\equiv \int_{0}^{\beta
}f_{o}^{1/2}\left( R\right) dt=f_{o}^{1/2}\left( R\right) \beta
=Cf_{i}^{1/2}\left( R\right) \beta $. Hence, the inverse
temperature is given by
\begin{equation}
\beta _{R}=2\pi xr_{+}^{2}\frac{R-3GM}{r_{+}-3GM}.  \label{betar}
\end{equation}
We shall also be interested in the Euclidean action, which turns
out to have a rather simple expression in terms of $r_+$. This is
derived in Appendix B, where it is shown that the difference in
Euclidean actions between the instanton and the background
solutions is given by
\begin{equation}
B=\frac{\pi }{GH_{o}^{2}}\left( 1-r_{+}^{2}H_{o}^{2}\right).
\label{B}
\end{equation}

\subsection{Some limiting cases}

Let us start with the case of low tension branes, $\sigma_k/M_p^2  \ll
H_{o}, H_o-H_i$. In this case the parameter $x$ is large compared
with $H_{o}$, $R\simeq x^{-1}$ is small, and $GMH_{o}\ll 1$. In
this limit the angle $\varphi$ in Eq.~(\ref{varphi}) is close to
$-\pi/2$ and $H_{o}r_{+}\simeq 1$. We have
\begin{equation}
B \simeq {2\pi \over H_o} \frac{16 \pi \sigma_k^3}{3 \epsilon^2},\quad\quad
\beta _{R} \simeq \frac{2\pi }{H_{o}}
.
\end{equation}
This is just the flat space expression for the energy of a
critical bubble, multiplied by Euclidean time periodicity of the
low curvature de Sitter space in which this bubble is embedded.

Next, we may consider the case of intermediate tension $H_o-H_i \ll \sigma_k/M_p^2 \ll H_o, H_i$. In this case, $x\approx H_i/\sqrt{2}$, $R\approx (\sqrt{3} x)^{-1}$, $H_o r_+\approx 1-GMH_o$, with $GMH_{o}\ll 1$, and we have
\begin{equation}
B\approx {16\pi^2\over 3\sqrt{3}}{\sigma_k\over H_o^3}, \quad\quad
\beta_R\approx {2\pi \over \sqrt{3}H_o}.
\end{equation}
In this case, the difference in pressure between inside and outside of the brane is insignificant compared with the brane tension term, which is balanced against collapse by the cosmological expansion.

Finally, in the limit of very large $\sigma_k $ we find that $3GM$ becomes
larger than $R$, namely, $3GM\rightarrow 4R/3$. This means that
$f_{o}\left( R\right) $ vanishes for some value $\sigma_k =\sigma
_{\max }$, given below in Eq. (\ref{sigmamax}), so it is not
sensible to consider the limit of very large $\sigma_k $ but just
the limit $\sigma_k \rightarrow \sigma _{\max }$. As we have
mentioned [see the discussion below Eq.~(\ref{fofactors})], the exterior metric in this
limit corresponds to the Nariai solution, with $ r_s=r_{+}=\left(
\sqrt{3}H_{o}\right) ^{-1}$. Replacing this value in (\ref{B}) we
find readily
\begin{equation}
B=\frac{2\pi }{3GH_{o}^{2}}.
\end{equation}
It is interesting to compare this value of $B$ with the corresponding one for the nucleation of black holes in the same de Sitter universe. This is described by the Nariai instanton \cite{perry}, which has the bounce action
\begin{equation}
B_N = \frac{\pi }{3GH_{o}^{2}}.
\end{equation}
Note that the difference $B-B_N = \pi/3GH_o^2 = A_{bh}/4G$, where $A_{bh}=4\pi r_s^2$ is the area of the black hole horizon in the Nariai solution. Hence, the probability of nucleating black holes divided by the probability of nucleating brane configurations characterized by the same mass parameter is just the exponential of the black hole entropy, as expected from general considerations (In this argument, we are of course neglecting the entropy stored in the field degrees of freedom living on the branes, which will only show up when the determinantal prefactor in the nucleation rate is evaluated).

Let us consider the value of $\beta _{R}$ in the limit $\sigma_k\to \sigma_{max}$. This is a singular limit in Eq. (\ref{betar}) due to the simultaneous vanishing of numerator and
denominator. Thus we will need to change to more appropriate
coordinates.
The fact that $r_s=r_{+}$ does not mean, though, that both
horizons coincide, since the coordinates $r,t$ become inadequate
in this case. Near this limit the metric outside takes the form
(\ref{metrico}), with
\begin{equation}
f_o(r)\approx A^{2}\left( 1-\frac{r}{r_{+}}\right) -\left(
1-\frac{r}{r_{+}} \right) ^{2},
\end{equation}
and $r\approx r_+$, plus higher orders in $A$. The constant $A$ is
the same parameter defined in (\ref{A}), but in this limit tends
to zero, $A^{2}=\sqrt{3}H_{o}\left( r_{+}-r_s\right) $. Now we
define new coordinates $\psi$ and $\lambda$ by
\begin{equation}
\cos \psi =1-\frac{2}{A^{2}}\left( 1-\frac{r}{r_{+}}\right) ,\quad
\lambda =\frac{A^{2}}{2}t,
\end{equation}
so that the metric becomes
\begin{equation}
ds^{2}=\sin ^{2}\psi\ d\lambda ^{2}+r_{+}^2 d\psi
^{2}+r_{+}^2d\Omega ^{2}.  \label{nariai}
\end{equation}
In these coordinates the cosmological horizon is at $\psi =0$ and
the black hole horizon is at $\psi =\pi $. Now in the limit
$A\rightarrow 0$ we just replace $r_{+}=\left(
\sqrt{3}H_{o}\right) ^{-1}$.

We must determine the position $\psi _{R}$ of the brane, which is
given as before by the matching conditions
(\ref{discont2}), where now the metric outside is
(\ref{nariai}). So, on the brane, we have
\begin{eqnarray}
ds_{\sigma_k }^{2} &=&\sin ^{2}\psi _{R}\ d\lambda
^{2}+r_{+}^{2}d\Omega ^{2} \\
&=&f_{i}\left( R\right) dt^{\prime 2}+R^{2}d\Omega ^{2}
\end{eqnarray}
and the extrinsic curvature on the outside of the brane is
$-(1/2)\partial _{\psi }g_{ab}$, with $g_{00}=\sin ^{2}\psi $ and $ g_{\Omega \Omega }=r_{+}^{2}$, i.e.,
$
K_{00} =-(1/r_{+})g_{00}\cot \psi ,
K_{\Omega \Omega } =0.$
The curvature inside is as before $K_{00}=g_{00}
\partial_r f_{i}^{1/2}$ and $K_{\Omega \Omega }=g_{\Omega
\Omega }f_{i}^{1/2}/r$, with $ f_{i}\left( r\right) =\left(
1-H_{i}^{2}r^{2}\right) $, so the Israel conditions give
\begin{eqnarray}
-\frac{1}{r_{+}}\cot\psi_R -\left(
f_{i}^{1/2}\right) ^{\prime }|_R &=&-4\pi G\sigma_k , \\
f_{i}^{1/2}\left( R\right) /R &=&4\pi G\sigma_k .
\end{eqnarray}
These equations are easily solved and give
\begin{eqnarray}
\sin \psi _{R}  &=&\left( \frac{
3H_{o}^{2}-H_{i}^{2}}{6H_{o}^{2}-H_{i}^{2}}\right) ^{1/2}, \\
\sigma_k  = \sigma_{\max} &=& 2M_{p}^{2}\sqrt{3H_{o}^{2}-H_{i}^{2}}
\label{sigmamax}
\end{eqnarray}
so $H_{i}$ must be less than $\sqrt{3}H_{o}$. Now regularity at
the cosmological horizon $\psi \simeq 0$ implies that $0\leq
\lambda /r_{+}\leq 2\pi $, so $\beta _{R}=\sin \left( \psi
_{R}\right) 2\pi r_{+}$. Hence,
\begin{equation}
\beta _{R}=\frac{2\pi }{\sqrt{3}H_{o}}\left(
\frac{3H_{o}^{2}-H_{i}^{2}}{6H_{o}^{2}-H_{i}^{2}}\right) ^{1/2}.
\end{equation}
Thus, also in this case, the effective temperature of the field
degrees of freedom living on the worldsheet will be of order
$H_0$ or higher.

\subsection{The prefactor for static instantons}

In flat space, and at finite temperature $T\gg
\epsilon_k/\sigma_k$, the instanton which is relevant for vacuum
decay is static and spherically symmetric in the spatial
directions. The fluctuations are periodic in Euclidean time, with
periodicity $\beta=1/T$. The worldsheet of the brane has the
topology $S^1\times S^2$, where the $S^1$ is the direction of
imaginary time, and the $S^2$ is the boundary of the ``critical"
bubble, a closed brane in unstable equilibrium between expansion
and collapse (this is in contrast with the zero temperature
instanton, where the world-sheet is a 3-sphere.) The radius of the
critical bubble is given by
$$
R_{\beta} = {2\sigma_k\over \epsilon_k},
$$
and the difference of the instanton action and the action for the
background is given by
$$
B_{\beta} = \beta E^{(0)},
$$
where $E^{(0)}= (4 \pi/ 3) \sigma_k R_{\beta}^2$ is the classical
energy of the critical bubble. The one loop quantum correction can
be written as (see e.g. \cite{fintemp})
\begin{equation}
W_{\Phi} = \beta F_\Phi= \beta (E_\Phi- T S_\Phi)
\label{staticweff}
\end{equation}
Here, $F_\Phi$ denotes the free energy and $E_\Phi$ is the
correction to the energy of the critical bubble due to the field
$\Phi$. This includes the zero point energy of $\Phi$ in the
presence of the bubble, as well as the thermal contributions.
Finally, $S_\Phi$ is the entropy. Thus, the nucleation rate takes
the form
$$
\Gamma_{\beta} = D e^{-B_\beta} \sim e^{-(B_\beta + k^2 W_\Phi)
}\sim  e^{-\beta F}\sim  e^{-E/T} e^S,
$$
where $E=E^{(0)}+ k^2 E_\Phi$ is the total energy, $F=E^{(0)}+k^2
F_\Phi$ is the total free energy and $S=k^2 S_\Phi$ is the total
entropy. Thus, for thermal instantons the determinant prefactor
does indeed include the exponential of the entropy. This is,
however, not the only role of the prefactor, since there is also
some correction to the energy of the bubble.

Consider, for simplicity a massless field $\Phi$. At sufficiently
high temperature $T \gg R_\beta^{-1}$, the entropy behaves as
$S_\Phi\propto T^2 R^2_\beta$. From $S_\Phi=-\partial
F_\Phi/\partial T$, it follows that $F_\Phi \approx -T\ S_\Phi
/3$. Hence
$$
\Gamma_{\beta} = D e^{-B_\beta} \sim e^{-B_\beta} e^{+ k^2
S_\Phi/3}.
$$
In this case, the prefactor clearly represents an ``entropy
enhancement". On the other hand, at lower temperatures, it is not
clear whether the prefactor represents an enhancement or a
suppression. At temperatures comparable to $R_\beta^{-1}$, the
vacuum energy term can be as important as the thermal
contributions, and the logarithm of the prefactor $D$ can have
either sign.

The case of the static instanton in de Sitter space is somewhat close to this low temperature situation, and without an explicit calculation it is not clear whether the prefactor represents an enhancement or a suppression of the nucleation rate. An interesting possibility would be that at sufficiently high Gibbons-Hawking temperature $T_o\sim H_o$, the thermal contribution may be sufficient to restore the symmetry, creating the desired attractive force amongst the branes. This is currently under research.

The static instanton presented in this Section may perhaps be  better suited to the scenario proposed by Feng et al. \cite{FMSW}, than the Coleman-de Luccia instanton. Ignoring the degeneracy factors, the action of the static instanton is always larger than the action of the CdL instanton. In this sense, it seems to correspond to a subdominant decay channel. However, as we have discussed in previous Sections, it might well be that the CdL instanton for multiple brane nucleation simply does not exist because of the repulsive force amongst the branes. This does not exclude the possibility that in the case of the static thermal instantons the symmetry is restored at high ambient (and worldsheet) dS temperature $\sim H_0$. In this situation, the decay through nucleation of coincident branes would only be possible through the static instanton \footnote{In the weak coupling limit, we have checked that indeed the $U(k)$ symmetry is not restored in the case of the CdL instanton \cite{gm}.}. At low $H_o$, the thermal contribution might not be sufficient to restore the symmety and stacks of branes may simply not hold together, destroying the possibility of further decay by coincident brane nucleation. Also, the prefactor and the Euclidean action have an exponential dependence on the ambient temperature $\sim H_o$, and can be much suppressed at the present epoch, contributing to the stability of the present vacuum (in contrast with the Coleman-de Luccia case).

\subsection{An entropy bound}

A potentially worrying aspect of coincident brane nucleation in the CdL case is whether the degeneracy factor may grow without bound as we increase the number of branes \cite{FMSW}.  As we have seen, this will not happen for the case of the static instanton discussed in this section, since nucleation of coincident branes cannot involve arbitrarily large $k$. Indeed, there is a maximum value of the combined tension of the branes
$\sigma_k\ll \sigma_{max}\sim M_p^2 H_o$, given in Eq. (\ref{sigmamax}), beyond which the instanton simply does not exist.
In this limit, the metric outside of the branes approaches the Nariai solution.

From this observation, we can easily derive a bound on the maximum entropy which can be stored in the stack of branes. Indeed, the static instanton represents a spherical bubble in unstable equilibrium between undeterred expansion or collapse into a black hole.
The entropy can only increase when the stack of branes collapses, and hence the entropy of the coincident branes cannot exceed the entropy of the Nariai black hole. The stack of branes with tension
$\sigma_k \to \sigma_{max}$ has the same radius as the horizon of the Nariai black hole, and so, the entropy of the stack of branes is bounded by one fourth of its own area, in natural units. A corollary is that the nucleation rate of coincident branes in the limit $\sigma_k \to \sigma_{max}$ is bounded above by the nucleation rate of Nariai black holes of the same mass.

\section{Conclusions and discussion}

In this paper we have investigated the possibility of coincident brane nucleation by a four-form field, in connection with string motivated scenarios for the neutralization of the effective cosmological constant.

In four dimensions, and after the moduli are stabilized, the branes repel each other at distances larger than the inverse mass of the moduli. At shorter distances, their interactions will be model dependent, but in the simplest models the branes do not attract at the classical level. In this situation, it is unclear whether the Coleman-de Luccia (CdL) instanton for nucleation of coincident branes really contributes to the semiclassical decay rate, since it would have too many zero modes and negative modes.

Assuming that the CdL instanton exists for the nucleation of coincident branes (that is, assuming an attractive short range interaction amongst the branes in the stack), we have investigated the degeneracy factor accompanying the formula for the nucleation rate, due to the large number of worldsheet degrees of freedom. We have modeled such degrees of freedom by a weakly coupled SYM $U(k)$ gauge theory, which is unbroken when the branes are coincident. We find that the degeneracy factor does not depend very strongly on the ambient de Sitter (dS) temperatures before or after the nucleation event. Rather, it depends only on the radius of the instanton. Hence the degeneracy factors can be quite important even when the ambient dS temperature is as low as it is today. This may indicate that nucleation of coincident branes via the CdL instanton is in fact impossible, otherwise the present vacuum would immediately decay.

If the CdL instanton for coincident branes does not exist, stacks of branes may still nucleate through a ``static" instanton which represents pair creation of critical bubbles, in unstable equilibrium between expansion and collapse. This is the analog of the instanton for thermal activation in flat space. Despite the absence of a classical attractive force, the branes could be held together by thermal corrections to the interbrane potential, which tend to favor the symmetric phase (where branes are on top of each other). The calculation of this thermal effective potential for the static instanton is currently under research. One may ask whether a similar symmetry restoration may not happen for the CdL instanton. In this case the calculation has been done in \cite{gm}, where it is shown that the one loop potential does not help restoring the symmetry. So it is conceivable that the branes may stick together for the static instanton but not for the CdL instanton, in which case the former would be the relevant decay channel.

To conclude, we have presented some evidence that the ``saltatory" relaxation scenario of \cite{FMSW} may be difficult to implement via the
CdL instanton, since saltation would be hard to stop at present. Rather, we have speculated that it may be easier to implement through the static instanton.
In the scenarios proposed in Ref. \cite{FMSW} for the saltatory relaxation of the cosmological constant, two different possibilities were suggested for the effective temperature of the worldsheet degrees of freedom, namely $T_1 \sim H_o$ and $T_2 \sim (H_o H_i)^{1/2}$, where $H_i$ and $H_o$ are the expansion rates before and after nucleation. We have shown that for the static instanton, the relevant temperature is comparable to the ambient de Sitter temperature $\sim H_o$ before the tunneling. Hence, 
the nucleation rate of coincident branes would be unsupressed at large ambient de Sitter temperature, but exponentially suppressed at present, which is of course desirable.

Clearly, many issues need to be addressed before a scenario based on coincident brane nucleation can be used to successfully explain the smallness of the observed cosmological constant.
A considerable advance would be to understand why the large $\Lambda_{eff}$ relaxes to the small $\Lambda_{obs}$ instead of plunging directly into deep AdS space (the latter jump involves a larger number of coincident branes and would be rewarded by a larger degeneracy factor). In Ref. \cite{FMSW} an explanation was offered, based on a ``uniquely weak" form of the anthropic principle.
As explained in Section II, any relaxation mechanism requires the gap $\Delta\Lambda$ in the discretuum of $\Lambda_{eff}$ not to be much larger than $\Lambda_{obs}$ (otherwise it becomes a problem to understand why, accidentally, there happens to be an allowed vacumm so close to zero, at $\Lambda_{eff}=\Lambda_{obs}\ll \Delta\Lambda$). In \cite{FMSW} it was proposed that $\Delta\Lambda = a \Lambda_{obs}$ with $a\sim 1$, saturating the above requirement. Then the allowed $\Lambda_{eff}$ would take values in the sequence $...,(1-a)\Lambda_{obs},\ \Lambda_{obs},\
(1+a)\Lambda_{obs},\ (1+2a)\Lambda_{obs},...$. If we start from a large $\Lambda_{eff}$, then the enhancement of brane nucleation for large $k$ favours a jump to the lowest value in the above list which is still compatible with the existence of observers. FMSW suggested that the value $(1-a)\Lambda_{obs}$ may already be too small for observers to emerge, making the vacuum with the value $\Lambda_{obs}$ the favourite destination.

Finally, one should try to embed this scenario in a cosmological context, taking into account the restrictions imposed by homogeneity and isotropy.
If unsuppressed saltation happened after inflation, then we would have seen signals of it in the microwave background. Indeed, bubbles which nucleate after thermalization are still rather small at the time of decoupling, and we would see different domains with different values of $\Lambda_{eff}$ separated by fast moving stacks of branes, which would presumably cause large perturbations in the gravitational potential. Hence, saltation should occur during inflation, and switch off somewhat before the end of it. This may impose certain constraints on the space of parameters such as the tension and charges of the branes, or alternatively, on the ambient temperature below which the instanton with coincident branes simply does not exist (e.g. because thermal symmetry restoration is no longer effective). Also, it should be clarified what might be the advantages of a saltatory ``neutralization" scenario over the ``randomization" scenarios discussed in Section II. A possible advantage is that saltatory relaxation operates very quickly, and hence it does not require eternal inflation to take place (as required in the randomization scenarios). A fuller discussion of these issues is left for further research.

\section*{Acknowledgements}

J.G. is grateful Roberto Emparan and Alex Vilenkin for very useful discussions.
A.M. was supported by a postdoctoral fellowship of the Ministerio
de Educaci\'on y Cultura, Spain. The work by J.G. was
supported by CICYT Research Projects FPA2002-3598, FPA2002-00748,
and DURSI 2001-SGR-0061.

\appendix

\section{Heavy fields on the branes}

The expression (\ref{finald}) for the nucleation rate is only valid for $m_- R \ll 1$. For completeness, here we shall discuss the limit $m_- R \gg 1$. In this case the heavy degrees of freedom decouple, and they simply contribute a finite renormalization of the parameters in front of different operators in the classical Lagrangian.

For scalar fields we have
\begin{equation}
W_- = -\frac{\pi}{6} m_-^3 R^3 + \frac{\pi}{4} m_- R + ...
\label{effacscalar}
\end{equation}
As discussed in Section VII, the first term in this expansion corresponds to a finite renormalization of the brane tension, multiplied by the worldvolume of the stack of branes. The second term correspond to a finite renormalization of the coefficient in front of the worldsheet Ricci scalar. This term was not present in the classical action we started with, but evidently it can be generated by quantum corrections. The scalar contribution (\ref{effacscalar}) tends to decrease the tension of the stack of branes. This tends to favour the nucleation rate at large $m_- R$, as represented in Fig. \ref{prefactor}. However, whether an actual enhancement really occurs will be model dependent, since all massive species, and not just the scalars, contribute finite renormalizations of the parameters in the action. In Section VII we have assumed that there is an attractive short range force amongst the branes, which we have modeled as a mass term for the scalars representing the relative positions of the branes. One may expect that the same mechanism which generates a potential for the scalars may generate masses also for their fermionic partners. From (\ref{wpsi}), heavy fermions give a contribution to the effective potential of the form
\begin{equation}
\frac{1}{2} W_\Psi = +\frac{\pi}{6} m_\Psi^3 R^3 + \frac{\pi}{8} m_\Psi R + ... \quad (m_\Psi R \gg 1)
\label{effacfermion}
\end{equation}
per degree of freedom. This gives a positive renormalization of the brane tension (which tends to suppress the nucleation rate). In the special case where $m_-=m_\Psi$, the brane tension does not renormalize, but each pair of fields will still contribute a finite renormalization of the subleading term $
\Delta W = (3\pi/ 8) m_- R
$
which would suppress the nucleation rate. Generically, however, scalars and fermions may wind up with different masses (since supersymmetry is broken at some level), and the tension will be renormalized. Dividing the leading term in $W$ by the worldvolume ${\rm Vol}[{S^3}]=2\pi^2 R^3$, each scalar and fermionic degree of freedom contributes a brane tension renormalization of the form
\begin{equation}
\Delta\sigma \approx \frac{m_\Psi^3 - m_-^3}{12\pi}.
\label{deltasigma}
\end{equation}
Similarly, there will be a renormalization of the induced Newton's constant $G_N$ on the worldsheet, of the order $\Delta G_N^{-1} \sim (m_\Psi - m_-)$ for each pair of heavy field species. These changes will modify the instanton solution. For $k$ coincident branes, the number of such fields grows as $k^2$, and the effect of these finite renormalizations can be quite dramatic. The nucleation rate will take the form
\begin{equation}
\Gamma_k \sim
{\sigma^2 R^2}
(A g^2 R)^{k^2/2} e^{-B_k^{\rm ren}} \quad (m_- R \gg 1).
\label{largemassrate}
\end{equation}
where $A \sim 1$ is a constant which depends on whether some (or all) of the fermionic species have decoupled or not. The bounce action $B_k^{ren}$ in the exponent is calculated by using the renormalized values of the parameters. If $\Delta\sigma < 0$, then the renormalized tension $\sigma_k \sim k \sigma + (k^2-1) \Delta\sigma$  sharply decreases for large $k$, leading to unsuppressed tunneling rate. On the contrary, for $\Delta\sigma>0$ the nucleation of coincident branes is strongly suppressed.

\section{Euclidean action for the static instanton}

The action is given by
\begin{equation}
S_{E}(I)=\sigma \int d^{3}\xi \sqrt{\gamma }+\int
d^{4}x\sqrt{g}\left( \rho _{V}-\frac{\mathcal{R}}{16\pi G}\right).
\end{equation}
On shell, the scalar curvature is given by,
\begin{equation}
\mathcal{R}\sqrt{g}=32\pi G\rho _{V}\sqrt{g}+24\pi G\sigma \int
d^{3}\xi \sqrt{\gamma }\delta ^{(4)}\left( x-x\left( \xi \right)
\right),
\end{equation}
and hence the instanton action is given by
\begin{equation}
S_{E}(I)=-\frac{\sigma }{2}\int d^{3}\xi \sqrt{\gamma }-\int
d^{4}x\rho _{V}\sqrt{g}.  \label{actionstatic}
\end{equation}
The first integral in (\ref{actionstatic}) is just the volume of a
two-sphere of radius $R$ times $\beta_R$. The second integral in
(\ref{actionstatic}) splits into two contributions from the two
regions,
\begin{eqnarray}
&&\rho _{i}\int_{0}^{R}Cdtdr4\pi r^{2}+\rho
_{o}\int_{R}^{r_{+}}dtdr4\pi
r^{2} \\
&=&\rho _{i}C\beta \frac{4}{3}\pi R^{3}+\rho _{o}\beta
\frac{4}{3}\pi \left( r_{+}^{3}-R^{3}\right)
\end{eqnarray}
So the instanton action is
\begin{equation}
S_{E}(I)=-2\pi R^{2}\sigma f_{o}^{1/2}\left( R\right) \beta
-R^{3}\frac{H_{i}^{2}}{ 2G}\frac{f_{o}^{1/2}\left( R\right)
}{f_{i}^{1/2}\left( R\right) }\beta -\left( r_{+}^{3}-R^{3}\right)
\frac{H_{o}^{2}}{2G}\beta .
\end{equation}
After some algebra $S_{E}(I)$ can be written in the simple form
\begin{equation}
S_{E}(I)=-\frac{\pi r_{+}^{2}}{G} .
\end{equation}
The exponent $B$ which gives the probability for brane nucleation
is  the difference in Euclidean actions between instanton and
background. The action of the background is just $ S_{E}= -\pi/
GH_{o}^{2}, $ so the difference in Euclidean actions leads to Eq.
(\ref{B}).


\begin{thebibliography}{99}

\bibitem{teitelboim} J. D. Brown and C. Teitelboim, Phys. Lett.
\textbf{B195}, 177, (1987); Nucl. Phys. \textbf{297},
787 (1988).

\bibitem{FMSW} J. L. Feng, J. March-Russell, S. Sethi and F. Wilczek,
Nucl. Phys. \textbf{B602}, 307 (2001).

\bibitem{deLuccia} S. Coleman and F. De Luccia, Phys. Rev. \textbf{D21},
3305 (1980).

\bibitem{solutions} J. Garriga and A. Vilenkin, Phys. Rev. \textbf{D64},
023517 (2001).

\bibitem{coleman} S. Coleman, Phys. Rev. \textbf{D15}, 2929 (1977).

\bibitem{alexgia} G. Dvali and A. Vilenkin, Phys. Rev. \textbf{D67},
046002 (2003).

\bibitem{alexgiahierarchy}
G.~Dvali and A.~Vilenkin,
arXiv:hep-th/0304043.

\bibitem{MSW} H.~Martel, P.~R.~Shapiro and S.~Weinberg, Ap.J
\textbf{492}, 29 (1998).

\bibitem{prelam} J. Garriga and A. Vilenkin, Phys. Rev. \textbf{D67}
043503 (2003).

\bibitem{mario} J. Garriga, M. Livio and A. Vilenkin,
Phys. Rev. \textbf{D61}, 023503 (2000)


\bibitem{bopo}
R.~Bousso and J.~Polchinski,
JHEP {\bf 0006} (2000) 006 [arXiv:hep-th/0004134].


\bibitem{vilenkin81}
A.~Vilenkin,
Phys.\ Rev.\ D {\bf 23} (1981) 852.

\bibitem{johnson}
C.~V.~Johnson, ``D-Branes,'' {\it  Cambridge, USA: Univ. Pr. (2003)
548 p}.

\bibitem{israel} W. Israel, Nuovo Cimento \textbf{44B}, 1 (1966), and
correction in \textbf{48B}, 463 (1967).

\bibitem{halliwell87}
J.~J.~Halliwell,
Phys.\ Lett.\ B {\bf 185} (1987) 341.

\bibitem{modstab}
S.~Kachru, R.~Kallosh, A.~Linde and S.~P.~Trivedi,
Phys.\ Rev.\ D {\bf 68}, 046005 (2003)
[arXiv:hep-th/0301240];
S.~Kachru, R.~Kallosh, A.~Linde, J.~Maldacena, L.~McAllister and S.~P.~Trivedi,
arXiv:hep-th/0308055.
\bibitem{Giddings:2003zw}
S.~B.~Giddings,
Phys.\ Rev.\ D {\bf 68} (2003) 026006
[arXiv:hep-th/0303031].


\bibitem{nogo}
J.~M.~Maldacena and C.~Nunez,
Int.\ J.\ Mod.\ Phys.\ A {\bf 16} (2001) 822 [arXiv:hep-th/0007018].

\bibitem{cawe}
P.~Candelas and S.~Weinberg,
Nucl.\ Phys.\ B {\bf 237} (1984) 397.

\bibitem{nemanja}
N.~Kaloper, J.~March-Russell, G.~D.~Starkman and M.~Trodden,
Phys.\ Rev.\ Lett.\  {\bf 85} (2000) 928
[arXiv:hep-ph/0002001].

\bibitem{casimir}
Y.~P.~Goncharov and A.~A.~Bytsenko,
Class.\ Quant.\ Grav.\  {\bf 8}, L211 (1991);
D.~Muller, H.~V.~Fagundes and R.~Opher,
Phys.\ Rev.\ D {\bf 66} (2002) 083507
[arXiv:gr-qc/0209103].

\bibitem{basu} R. Basu, A. Guth and A. Vilenkin, Phys. Rev. \textbf{D44},
340 (1991).

\bibitem{nucrates}  J.~Garriga,
Phys.\ Rev.\ D {\bf 49} (1994) 6327. [arXiv:hep-ph/9308280].

\bibitem{gm} J. Garriga and A. Megevand, in preparation.



\bibitem{covariant} J.~Garriga and A.~Vilenkin,
Phys.\ Rev.\ D {\bf 44} (1991) 1007;
Phys.\ Rev.\ D {\bf 45} (1992) 3469;
J.~Guven,
Phys.\ Rev.\ D {\bf 48} (1993) 4604
[arXiv:gr-qc/9304032].

\bibitem{vachalex}
T.~Vachaspati and A.~Vilenkin,
Phys.\ Rev.\ D {\bf 43} (1991) 3846.

\bibitem{montes}
X.~Montes,
Int.\ J.\ Theor.\ Phys.\  {\bf 38} (1999) 3091
[arXiv:gr-qc/9904022].

\bibitem{bkk}
V.~A.~Berezin, V.~A.~Kuzmin and
I.~I.~Tkachev,
Phys.\ Scripta {\bf T36} (1991) 269.

\bibitem{hiscock}
D.~A.~Samuel and W.~A.~Hiscock,
Phys.\ Rev.\ D {\bf 44} (1991) 3052.

\bibitem{perry}
P.~Ginsparg and M.~J.~Perry,
Nucl.\ Phys.\ B {\bf 222} (1983) 245.

\bibitem{nariai} H. Nariai, Science Reports of the Tohoku Univ. 35 (1951) 62.

\bibitem{fintemp} A.~D.~Linde,
Nucl.\ Phys.\ B {\bf 216} (1983) 421
[Erratum-ibid.\ B {\bf 223} (1983) 544].


\end{thebibliography}
\end{document}